\pdfoutput=1

\documentclass[dvips,11pt]{article}

\usepackage{subfigure}
\usepackage[pdftex]{graphicx}
\usepackage{url}
\usepackage{amsmath}

\usepackage[a4paper,bindingoffset=0.2in,%
            left=0.7in,right=0.7in,top=0.6in,bottom=1.5in,%
            footskip=0.5in]{geometry}
\usepackage{blindtext}
\usepackage{setspace}
\begin{document}

\title{Ferromagnetic layer thickness dependence of the Dzyaloshinskii-Moriya interaction and spin-orbit torques in Pt$\backslash$Co$\backslash$AlO$_{x}$}

\author{R. Lo Conte$^{1,2,\ast}$, G. V. Karnad$^{1,\ast}$, E. Martinez$^{3}$, K. Lee$^{1}$, N.-H. Kim$^{4}$, D.-S. Han$^{5}$, J.-S. Kim$^{6}$,\\ S. Prenzel$^{1}$, T. Schulz$^{1}$, C.-Y. You$^{4}$, H. J. M. Swagten$^{5}$, M. Kl{\"a}ui$^{1,2,\ast\ast}$}

\date{\today}

\maketitle




\section*{}
$^{1}$Johannes Gutenberg-Universit{\"a}t, Institut f{\"u}r Physik, Staudinger Weg 7, 55128 Mainz, Germany\\
$^{2}$Graduate School of Excellence \textquotedblleft{Materials Science in Mainz}\textquotedblright (MAINZ), Staudinger Weg 9, 55128 Mainz, Germany\\
$^{3}$Departamento Fisica Applicada, Universidad de Salamanca, plaza de los Caidos s/n E-38008, Salamanca, Spain\\
$^{4}$Department of Emerging Materials Science, DGIST, Daegu 42988, Republic of Korea\\
$^{5}$Department of Applied Physics, Center for NanoMaterials, Eindhoven University of Technology, PO Box 513, 5600 MB Eindhoven, The Netherlands\\
$^{6}$Research Center for Emerging Materials, DGIST, Daegu 42988, Republic of Korea\\
$^{\ast}$These authors contributed equally\\
\textit{$^{\ast\ast}$} klaeui@uni-mainz.de
\section*{Abstract}
We report the thickness dependence of Dzyaloshinskii-Moriya interaction (DMI) and spin-orbit torques (SOTs) in Pt$\backslash$Co(t)$\backslash$AlO$_{x}$, studied by current-induced domain wall (DW) motion and second-harmonic experiments. From the DW motion study, a monotonous decay of the effective DMI strength with an increasing Co thickness is observed, in agreement with a DMI originating at the Pt$\backslash$Co interface. The study of the ferromagnetic thickness dependence of spin-orbit torques reveals a more complex behavior. The effective SOT-field driving the DW motion is found to initially increase and then saturate with an increasing ferromagnetic thickness, while the effective SOT-fields acting on a saturated magnetic state exhibit a non-monotonic behavior with increasing Co-thickness. The observed thickness dependence suggests the spin-Hall effect in Pt as the main origin of the SOTs, with the measured SOT amplitudes resulting from the interplay between the varying thickness and the transverse spin diffusion length of the Co layer.     
\section{Introduction}
The possibility of manipulating magnetization by spin-currents in a very efficient way is a key requirement for the design of novel spintronic devices \cite{parkin2015memory,fert2013skyrmions}, which promise to change the way digital information is processed and stored. In particular, the advantageous scaling of current-induced spin manipulation compared to the Oersted field-induced switching allows for lower power operation at small design rules. The driving mechanism behind current-induced magnetization dynamics pioneered for use in metallic ferromagnets in the last twenty five years has been the spin-transfer torque \cite{ralph2008spin,boulle2011current}. However, recently a novel and possibly more efficient approach to current-driven magnetization manipulation has been developed. In particular, very efficient current-induced spin dynamics has been observed in multilayer systems with an ultra-thin ferromagnetic layer sandwiched between two different non magnetic materials \cite{miron2011perpendicular,miron2011fast,garello2013symmetry,emori2013current,ryu2014chiral,torrejon2014interface,loconte2015PRB}. These new current-induced torques originate from spin-orbit effects (at least one of the two non-magnetic layers consists of an heavy metal) and so they are referred to as spin-orbit torques (SOTs) \cite{brataas2014spin}. Two different origins have been suggested for the SOTs. One is the spin accumulation induced at the [heavy metal]$\backslash$ferromagnet interface due to the bulk spin-Hall effect (SHE) in the heavy metal layer \cite{emori2013current,ryu2014chiral,sinova2015spin}. A second possible origin of the SOTs is the inverse spin-galvanic effect (ISGE) \cite{edelstein1990spin,ganichev2002spin}, which generates a non-equilibrium spin-density at both the top and the bottom interfaces of the ferromagnet. Both effects are expected to induce an effective torque whose strength is a function of the ferromagnetic layer thickness.\\In the same kind of materials stacks the presence of topologically non-trivial spin textures, such as homo-chiral domain walls has also been observed \cite{miron2011fast,emori2013current,ryu2014chiral,torrejon2014interface,loconte2015PRB}. Chiral domain walls are reported to be very stable against annihilation \cite{benitez2015magnetic} and to exhibit interesting transport properties \cite{miron2011fast,emori2013current,ryu2014chiral}, thus being very promising for technological applications. The origin of these chiral spin structures is the interfacial Dzyaloshinskii-Moriya interaction (DMI) \cite{dzyaloshinsky1958thermodynamic,moriya1960new,moriya1960anisotropic,crepieux1998dzyaloshinsky,thiaville2012dynamics,kashid2014dzyaloshinskii,yang2015anatomy}. The DMI is expected to primarily originate from the interface between the heavy metal and the ferromagnet, predicting a direct scaling of its effective strength with the inverse of the ferromagnetic layer thickness \cite{kashid2014dzyaloshinskii,yang2015anatomy,nembach2015linear}. Accordingly, both the DMI and the SOTs are expected to depend strongly on the materials system as well as on the layers thickness. So, the ferromagnetic layer thickness dependence of DMI and SOTs is directly linked to their origin. A pure interface-like effect on one side and a more complicated mechanism which includes bulk-like processes on the other are expected to scale differently with the thickness of the ferromagnetic layer \cite{kim2013layer,fan2014quantifying}. In particular, there are predictions that the DMI and the ISGE can have a common origin, which should result in a similar dependence on the thickness \cite{freimuth2014berry}. So, studying the sign and amplitude of DMI and SOTs as a function of the ferromagnetic layer thicknesses is a key necessity to reveal their origins.\\Materials stacks with a strong DMI as well as large SOTs are particularly promising, due to their rich physics as well as the possibility to be used in novel spintronic devices \cite{parkin2015memory}. One of the most promising of those materials stacks is the trilayer Pt$\backslash$Co$\backslash$AlO$_{x}$ \cite{miron2011perpendicular,miron2011fast,garello2013symmetry,belmeguenai2015interfacial,cho2015thickness}. Several manuscripts have reported the characterization of the DMI \cite{belmeguenai2015interfacial,cho2015thickness,kim2015improvement,han2016asymmetric} and of the SOTs \cite{miron2011perpendicular,garello2013symmetry,liu2012current} in this system. However, reports from different groups for nominally identical samples have often shown contradicting results possibly originating from different thicknesses and growth conditions and no systematic study has been provided to date. In particular, varying the Co thickness is a key challenge as it potentially allows to distinguish between effects from the interface with the Pt and the AlO$_{x}$ and from the bulk of the materials. Accordingly, a systematic and combined study of both the DMI and SOTs as a function of Co thickness is the key step needed to understand and tailor the spin-orbit effects in this system.\\Here we report on the characterization of the DMI as well as the SOTs in identical samples of Pt$\backslash$Co$\backslash$AlO$_{x}$. We extract the sign and the magnitude of DMI and SOTs as a function of the Co layer thickness combining two key techniques, namely: current-induced DW motion (CIDWM) and second harmonic Hall measurements. A detailed study of CIDWM in magnetic tracks is presented, which allows us to characterize the thickness dependence of the DMI and of the effective torque driving the DW motion. This is complemented by an in depth characterization of the SOTs employing second harmonic measurements in a Hall bar geometry. Comparing the thickness dependence of the DMI and the torques allows us to draw conclusions about their origin.\\In Sec. 2 we describe the experimental techniques and the corresponding set-ups used in our study. In Sec. 3 we report the characterization of current-induced DW motion in magnetic nanowires. Sec. 4 reports the DMI values extracted from the analysis of this DW motion data. Sec. 5 presents the measured thickness dependence of SOTs. In Sec. 6 we present a discussion of the main results, where we  discuss the ferromagnetic thickness dependence of the DMI and the SOTs and compare to literature.
\section{Sample and experimental set-up}
\begin{figure}[htbp]
	\centering
	  \subfigure[]
		  {\includegraphics[width=70mm]{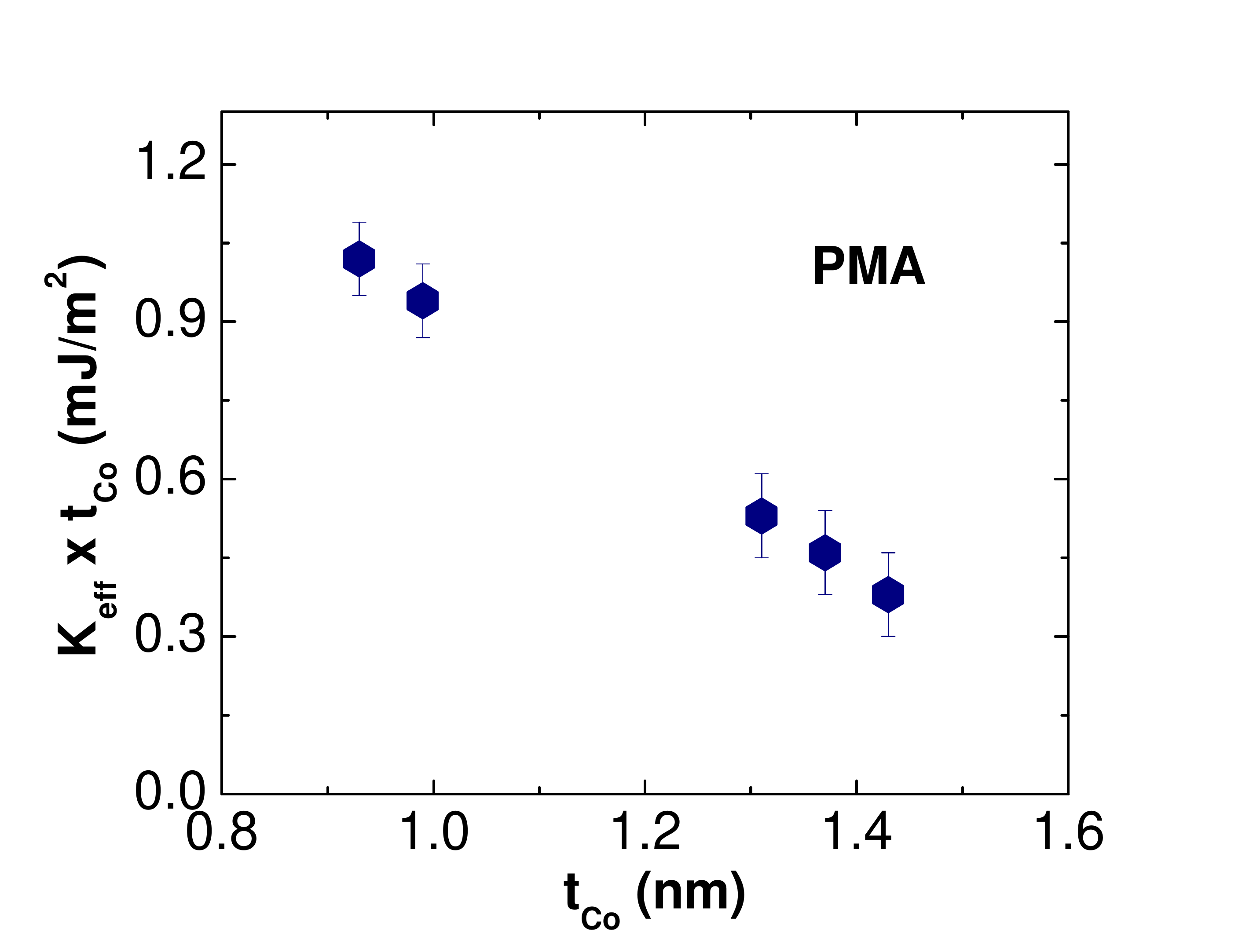}
		     \label{fig_K_eff}}
		\subfigure[]
		  {\includegraphics[width=70mm]{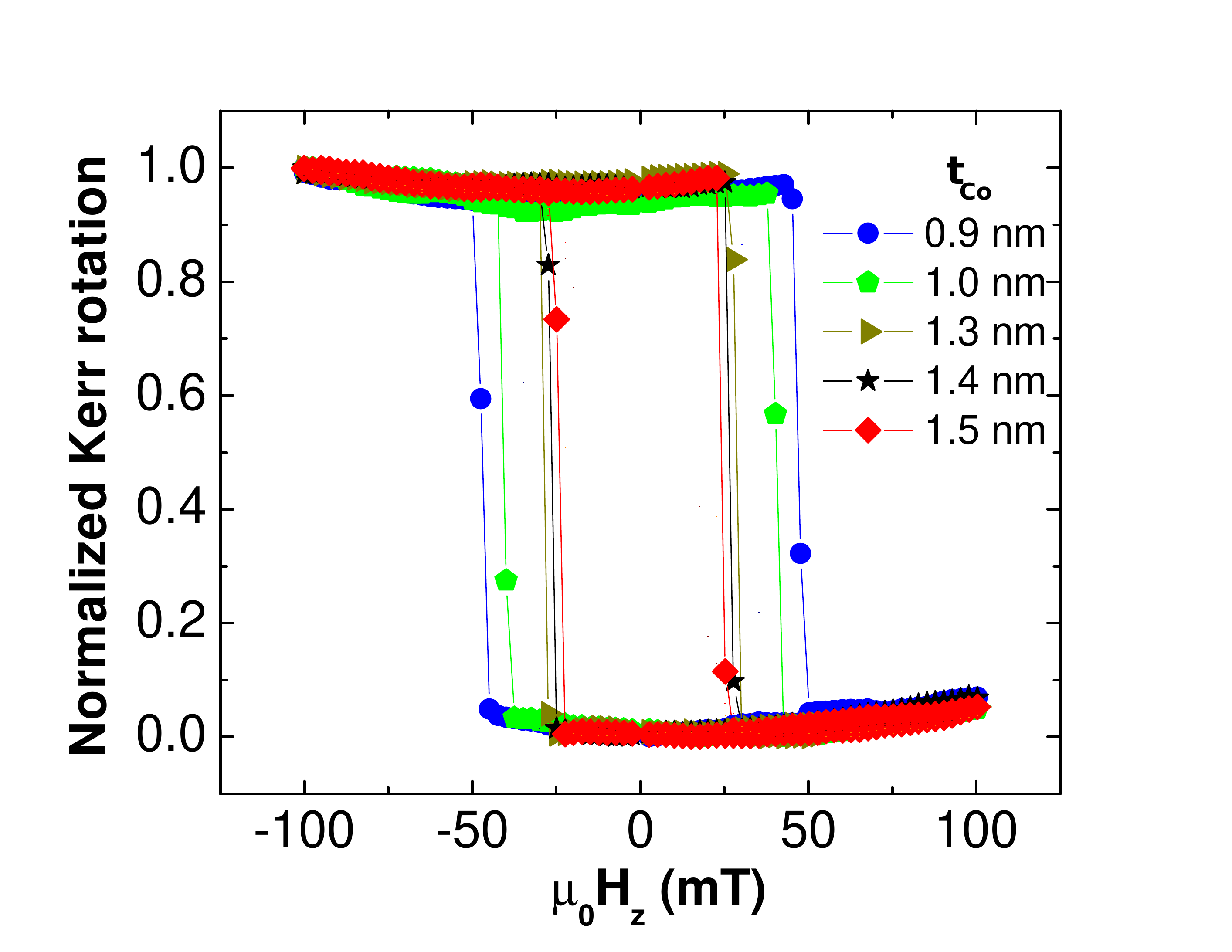}
		     \label{fig_MOKE_PtCoAlOx}}
	\caption{Magnetic characterization of the materials stack. (a) Effective magnetic anisotropy as a function of the Co-layer thickness. All the investigated devices exhibit PMA. The $K_{eff}$ values are calculated by using the $K_{i}$ and $M_{s}$ values obtained by BLS measurements (see Kim et al. \cite{kim2015improvement} for more details). (b) Polar-MOKE hysteresis loops for the magnetic multilayer. The external magnetic field is applied perpendicular to the sample surface, with the MOKE system operating in the polar configuration. The measurements are carried out at room temperature ($T=300$~K), probing 5 areas on the sample's surface with different thicknesses. The coercive field is observed to decrease with increasing $t_{Co}$.}
	\label{fig_characterization}
\end{figure}

\begin{figure}[htbp]
	\centering
	  \subfigure[]
		  {\includegraphics[width=80mm]{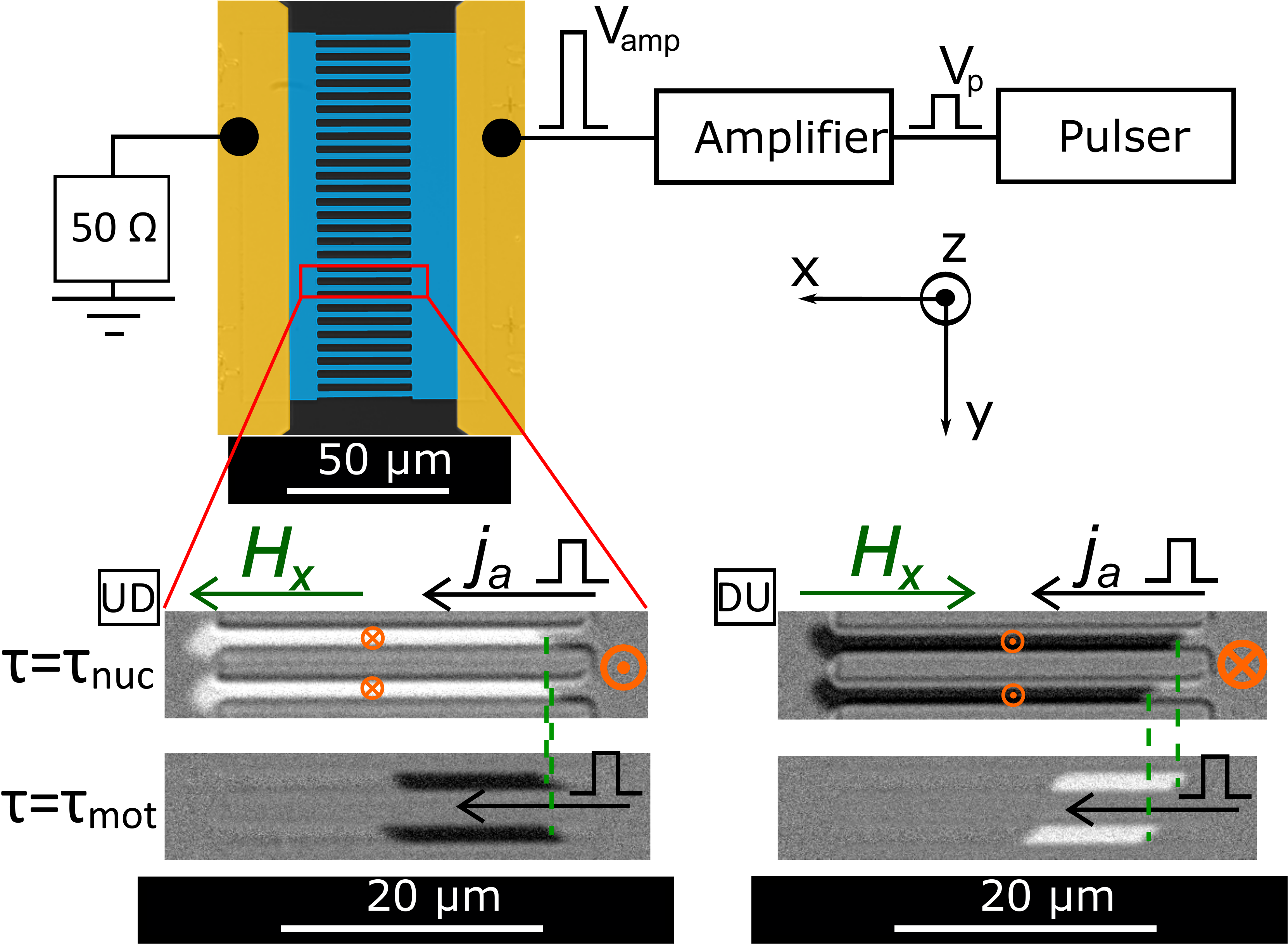}
		     \label{fig_CIDWM_protocol}}
		\subfigure[]
		  {\includegraphics[width=80mm]{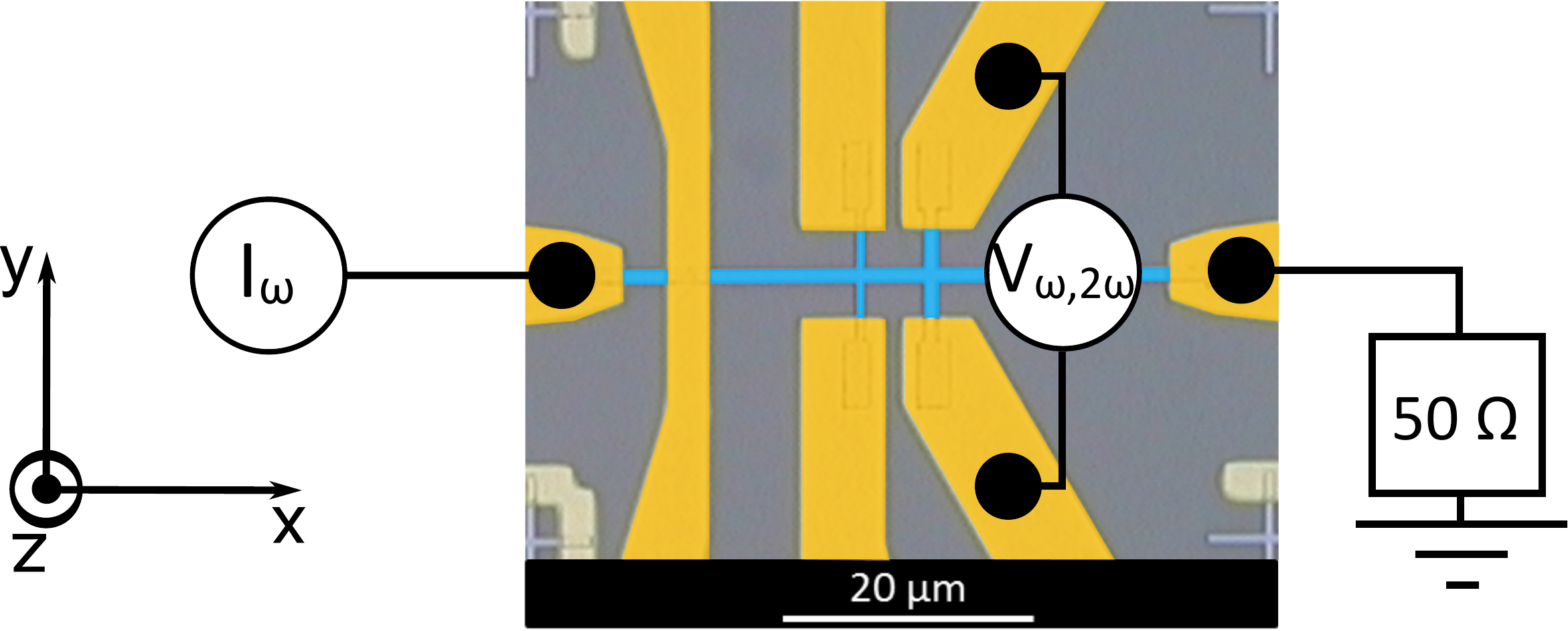}
		     \label{fig_SH_setup}}
	\caption{Experimental setups. (a) Schematic of the experimental setup for CIDWM measurements, including an optical image in false color (blue indicates the magnetic multilayer, yellow the Au contacts) of one of the devices used during the experiment, together with a schematic description of the typical experiment protocol. At $\tau=\tau_{nuc}$ the DWs are nucleated by the application of a fixed external magnetic field (50-100~mT) and current pulse ($8\times10^{11}$~A/m$^{2}$, 50~ns), in a device where the magnetization was pre-initialized by an external out-of-plain magnetic field. At $\tau=\tau_{mot}$ the DWs are moved by the injection of a train of current pulses. The DW displacements visible in the differential Kerr images are measured to calculate the DW velocity. The orange symbols represent the magnetization state in the magnetic device, while the green dashed lines indicate the initial position of the nucleated DWs. The procedure is shown for $\uparrow\downarrow$-DWs (left side) and $\downarrow\uparrow$-DWs (right side). (b) Schematic of the experimental setup for $2\omega$ measurements, including the false colors image of one of the Hall crosses used during the experiments. In the images, the yellow areas are the Au-contacts, the light blue indicates the magnetic devices.}
	\label{fig_setups}
\end{figure}

The material system that is employed for the patterning of the magnetic devices is the multilayer: Ta(4.0)$\backslash$Pt(4.0)$\backslash$Co(0.8--1.8)$\backslash$AlO$_{x}$(2.0) (all thicknesses in nm). The stack was deposited by magnetron sputtering technique on a Si$\backslash$SiO$_{2}$ substrate. The magnetic layer consists of a wedged Co layer, with a nominal increase in thickness of 1~nm over a length of 2~cm in one of the two in-plane directions ($x$-direction in the following). The wedged Co layer is obtained by an \emph{in-situ} moving shadow mask \cite{cho2015thickness}.\\Before patterning, the material stack is characterized by Brillouin light scattering (BLS) technique \cite{kim2015improvement} and magneto-optic Kerr effect (MOKE) magnetometry. The stack exhibits perpendicular magnetic anisotropy (PMA), with a spontaneous magnetization pointing along the out-of-plane (OOP) direction (see Fig. \ref{fig_characterization}). The observed PMA is induced by a strong interface anisotropy at the Pt$\backslash$Co interface \cite{nakajima1998perpendicular}. The effective anisotropy energy density, $K_{eff}=\frac{K_{i}}{t_{Co}}-\frac{1}{2}\mu_{0}M^{2}_{s}$, is found to decrease with increasing Co thickness (see Fig. \ref{fig_K_eff}), consistent with the presence of an interface anisotropy of $K_{i}=2.20\pm0.06$~mJ/m$^{2}$ and a saturation magnetization of $M_{s}=(1.42\pm0.02)\times10^{6}$~A/m, both obtained by BLS measurements \cite{kim2015improvement}. This is consistent with the results of magneto-optic Kerr effect (MOKE)-magnetometry measurements shown in Fig. \ref{fig_MOKE_PtCoAlOx}. Hysteresis loops for an applied OOP magnetic field, $\mathbf{H_{z}}$, are measured at different positions on the surface of the material stack with different $t_{Co}$. The coercive field is observed to decrease with increasing $t_{Co}$, in agreement with previous observations \cite{belmeguenai2015interfacial}. The loss of PMA with increasing $t_{Co}$ explains the decreasing coercive field as a function of an increasing Co thickness visible in Fig. \ref{fig_MOKE_PtCoAlOx}.
After magnetic characterization, the wedge sample is patterned into several devices with different thicknesses of the Co layer. The patterned devices consist of an array of several nanowires (NWs) in parallel (1.5-2.0~${\mu}$m in width and 25-28~${\mu}$m in length, see Fig. \ref{fig_CIDWM_protocol}) used for current-induced domain wall motion (CIDWM) experiments, and in Hall-crosses (1-2~${\mu}$m in width and 50~${\mu}$m in length, see Fig. \ref{fig_SH_setup}) used for the measurements of effective spin-orbit fields by the second harmonic ($2\omega$) technique \cite{garello2013symmetry,emori2013current}. The devices are patterned by electron-beam lithography and Ar-ion milling, at different positions on the sample surface corresponding to different $t_{Co}$.\\In both experimental setups a 50~$\Omega$-resistor is used to terminate the circuit to ground. An oscilloscope is used for measuring the pulse waveform, across its 50~$\Omega$-internal resistance, $R_{o}$. The total current flowing through the device is obtained by measuring the voltage across $R_{o}$. For the evaluation of the current density, $j_{a}$, the nominal thicknesses of the layers are used. In these type of thin film systems, the resistivity of the Ta layer is known to be around 4-5 times larger than is for Pt, while the Co layer and the Pt layer have a similar resistivity value \cite{emori2013current}. Accordingly, the calculated current densities are obtained considering the 4~nm-thick Ta bottom layer equivalent to a 1~nm-thick Pt layer. The conventional current density $\mathbf{j}_{a}$ is taken to be positive when it flows in the $+x$-direction (see Fig. \ref{fig_setups}), corresponding to an electron current density $\mathbf{j}_{e}$ flowing in the $-x$-direction.
\\Concerning the CIDWM experiment, the magnetic configuration of the wires is imaged by a wide field Kerr microscope in the polar configuration \cite{loconte2015PRB}. A magnetic coil is used for the generation of an external in-plane magnetic field. The experiments are carried out at T=300~K. The second harmonic measurements are carried out in Hall-crosses (see Fig. \ref{fig_SH_setup}). A small-amplitude sinusoidal ac current is applied with a frequency ($\omega/2\pi$) of 13.7 Hz. This induces periodic oscillations of the magnetization about its equilibrium direction. These periodic oscillations can be attributed to the effective fields generated by the injected current. The periodic oscillation of the magnetization results in an oscillation of the Hall resistance. The resulting Hall voltage has a second harmonic component ($2\omega$) that relates directly to the current-induced fields \cite{garello2013symmetry,pi2010tilting,hayashi2014quantitative}. The measurement is performed in two schemes: longitudinal (x-axis) and transverse (y-axis). In both schemes, the Hall voltage is measured during a sweep of the in-plane magnetic field (-400~mT to +400~mT). In the longitudinal (transverse) scheme the direction of the magnetic field is applied along (perpendicular to) the direction of the injected current. It is important to note that the Hall voltage also includes contributions from the planar Hall effect \cite{hayashi2014quantitative}, Nernst effect and Joule heating \cite{lee2014spin}. These effects are taken into account to extract artifact-free current induced effective fields. 
\section{Moving chiral domain walls}
The first type of experiment reported here is the study of current-induced DW motion (CIDWM). This allows us to establish if either the standard spin-transfer torque or the SOTs are the main driving mechanisms behind DW motion in this materials system and gauge the strength. Furthermore, the DW motion study allows one to determine if chiral DWs are present in the system, to extract their chirality and finally to obtain the sign and magnitude of the DMI for each investigated device.
\subsection{Current-induced domain wall motion}
Current-induced DW motion experiments are carried out in four different devices. The nominal thickness of the Co layer in the different devices is: 0.93~nm, 1.31~nm, 1.37~nm and 1.43~nm. The measurement protocol is described in Fig. \ref{fig_CIDWM_protocol}.
\begin{figure*}[htbp]
	\centering
	  \subfigure[]
		  {\includegraphics[width=75mm]{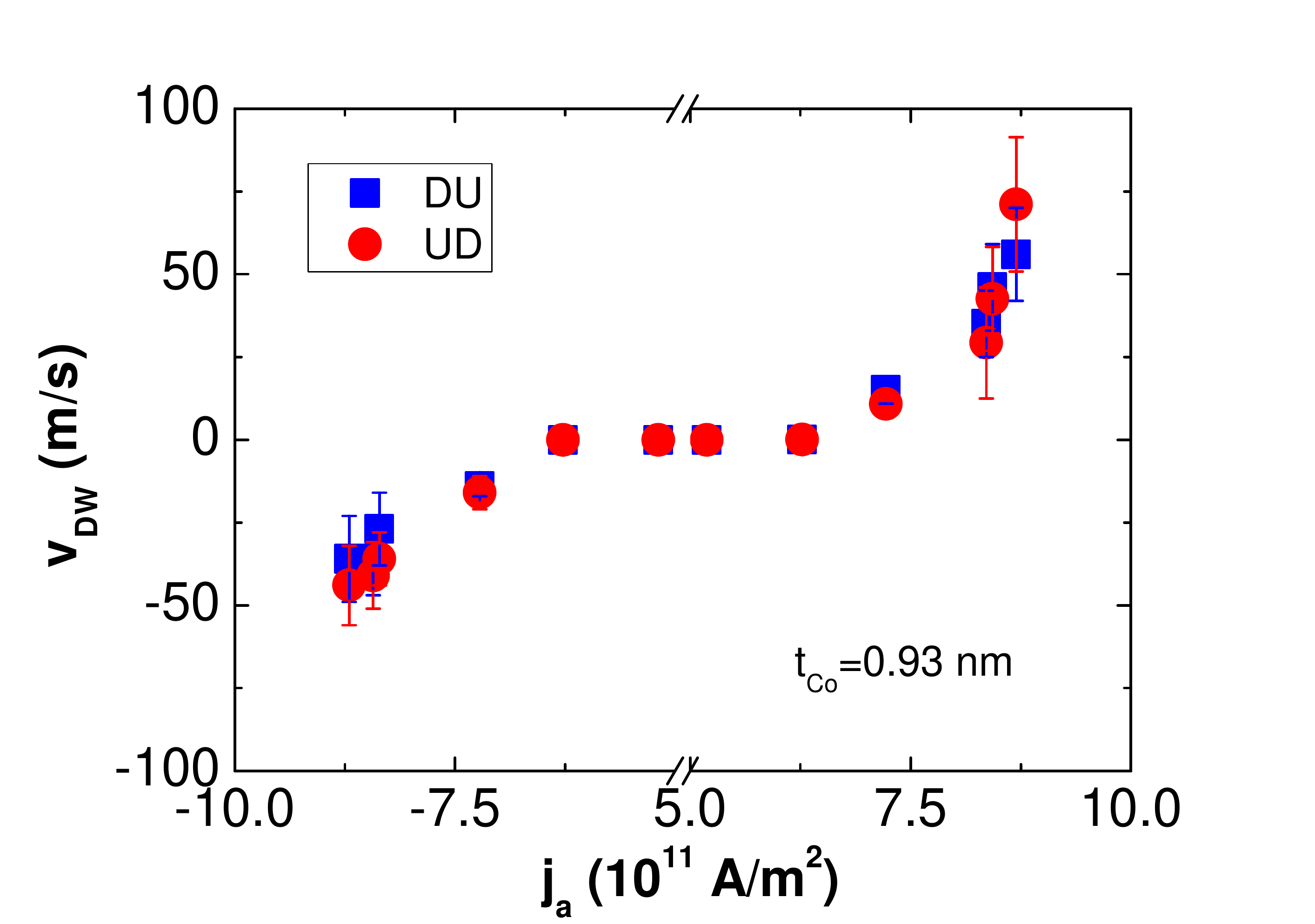}
		     \label{fig_CIDWM_1}}
		\subfigure[]
		  {\includegraphics[width=75mm]{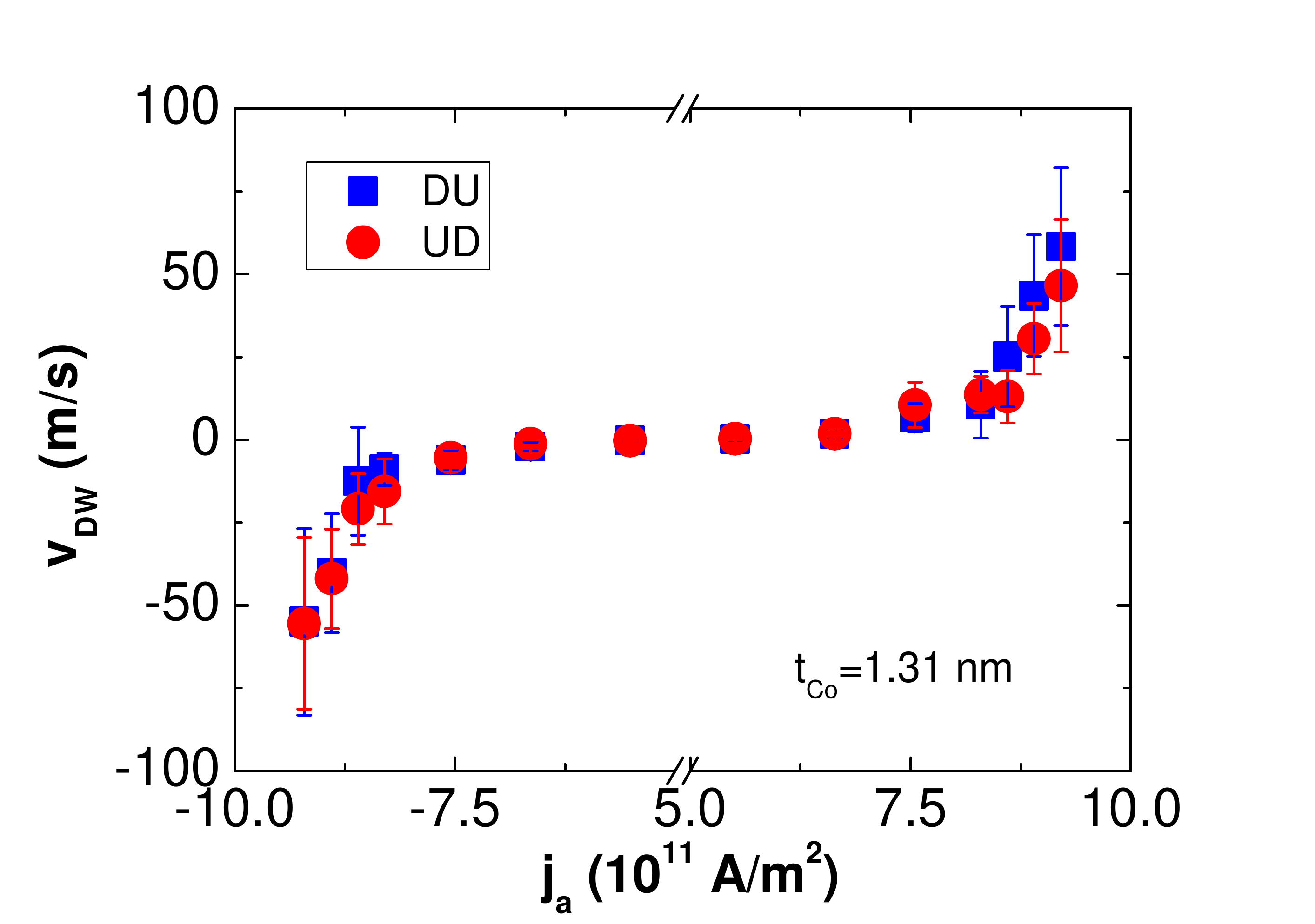}
		     \label{fig_CIDWM_3}}
		\subfigure[]
		  {\includegraphics[width=75mm]{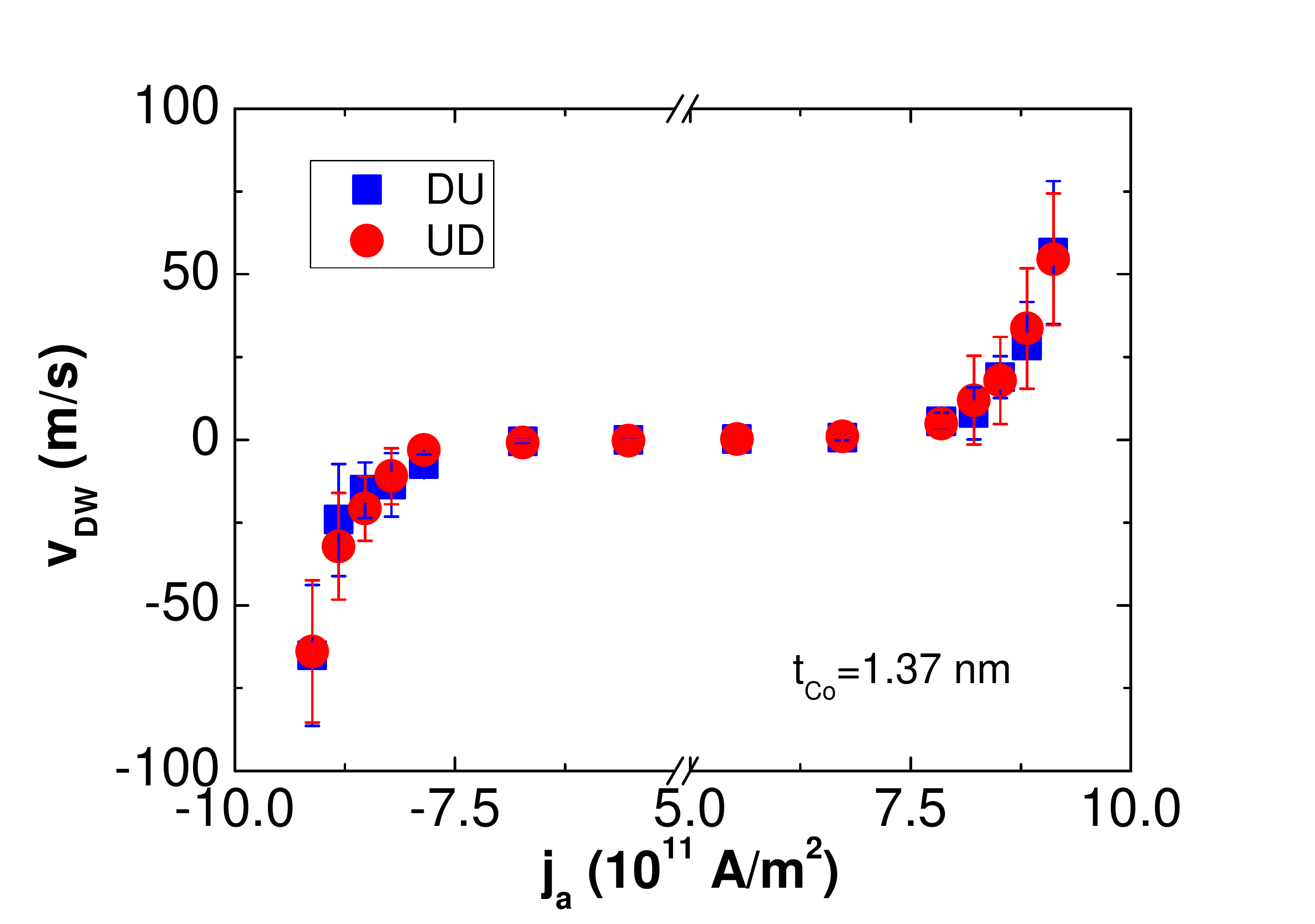}
		     \label{fig_CIDWM_4}}
		\subfigure[]
		  {\includegraphics[width=75mm]{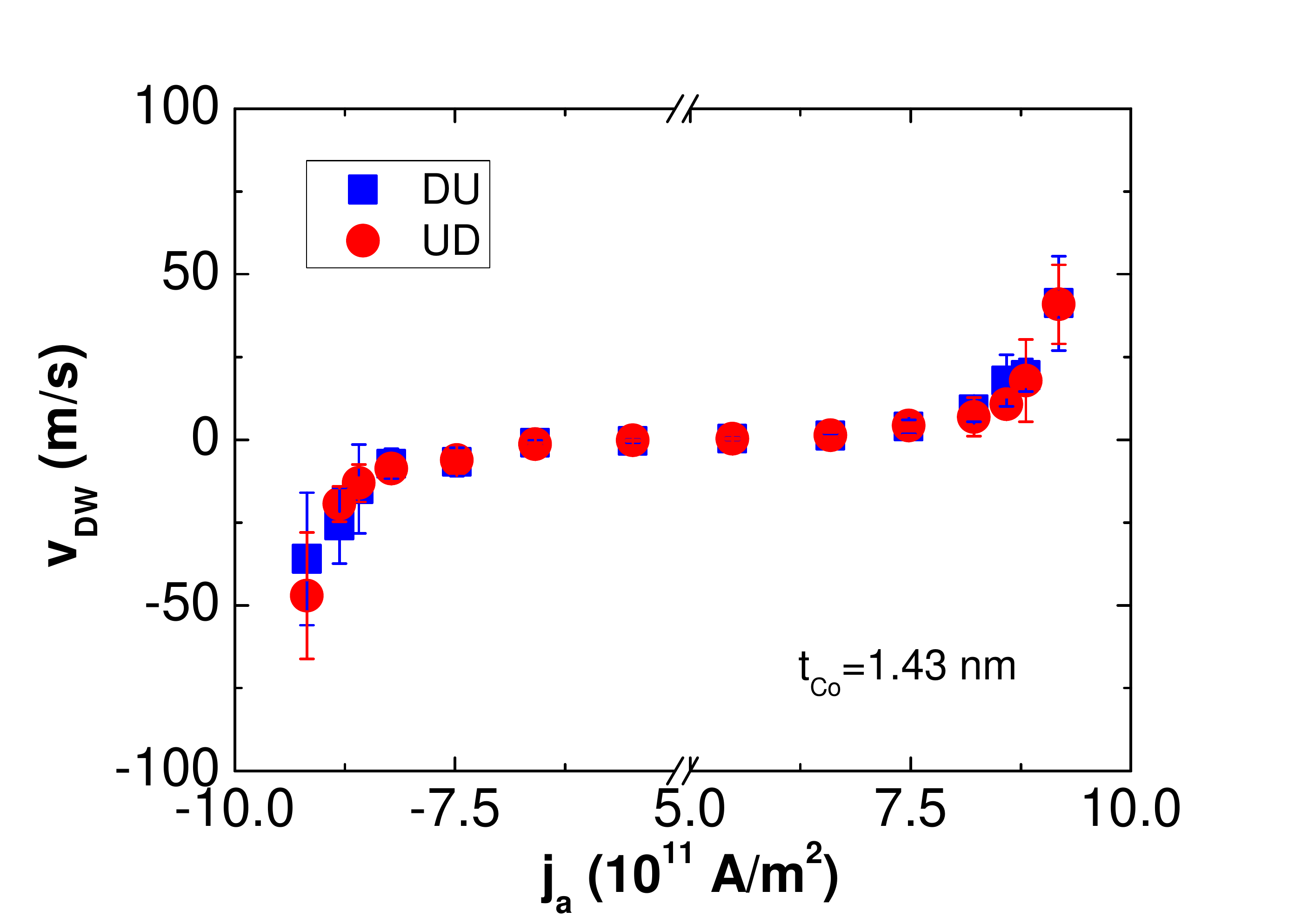}
		     \label{fig_CIDWM_5}}				
	\caption{Average DW velocity, $v_{DW}$, as a function of the current density, $j_{a}$, injected in the magnetic wires, for the 4 different devices. The DW velocity is reported for both $\uparrow\downarrow$-DWs (red dots) and $\downarrow\uparrow$-DWs (blue squares). The average velocities and the error bars (standard deviations) are calculated from several DW motion events, for each current density. (a) CIDWM data for $t_{Co}=0.93$~nm. (b) CIDWM data for $t_{Co}=1.31$~nm. (c) CIDWM data for $t_{Co}=1.37$~nm. (d) CIDWM data for $t_{Co}=1.43$~nm.}
	\label{fig_vDW-ja}
\end{figure*}
First, the magnetic wires are saturated in the \textquotedblleft{up}\textquotedblright ($\uparrow$, $+z$) or \textquotedblleft{down}\textquotedblright ($\downarrow$, $-z$) magnetization state by an external out-of-plane magnetic field. Second, the magnetization is reversed in all the NWs by the switching process presented in previous papers \cite{miron2011perpendicular,avci2012magnetization,loconte2014APL}, where each combination (parallel/anti-parallel) of field, $H_{x}$, and current, $j_{a}$, corresponds to a specific final state of the magnetization in the NWs. As a result, a DW ready to be displaced is obtained in each NW. The differential Kerr microscopy images at the bottom left (right) of Fig. \ref{fig_CIDWM_protocol} show the CIDWM experiment for $\uparrow\downarrow$- ($\downarrow\uparrow$-) DWs. At $\tau=\tau_{nuc}$ DWs are nucleated. At $\tau=\tau_{mot}$ the same DWs are moved by the injection of a train of current pulses.\\The average DW velocity is extracted as the ratio between the total displacement, $\Delta{x}$, visible in the differential Kerr microscopy images and the total pulsing time, $\Delta{\tau_{tot}}$. The total pulsing time is given by the total number of pulses, $n$, times the single pulse length, $\Delta{\tau_{p}}$ (full width at half maximum, 10-15~ns): $\Delta{\tau_{tot}}=n\Delta{\tau_{p}}$. Two consecutive current pulses are separated by 1~ms, in order to fully magnetically relax the domain wall in between pulses to have reproducible conditions. The average DW velocity, $v_{DW}$, as a function of the current density, $j_{a}$, is reported in Fig. \ref{fig_vDW-ja}, for the four different devices. The measurement is carried out for both $\uparrow\downarrow$-DWs (red dots) and $\downarrow\uparrow$-DWs (blue dots), for positive and negative $j_{a}$.\\In all the four devices, $\uparrow\downarrow$- and $\downarrow\uparrow$-DWs are observed to move in the same direction of the conventional current, $\mathbf{j}_{a}$, at approximately the same speed (within the error bars). The critical current density for inducing significant DW motion in each device is observed to be in the range $j_{a}\approx6.0-7.0\times10^{11}$~A/m$^{2}$, in line with what has been reported in previous works on the same materials system \cite{miron2011fast,moore2008high}. From the result that both domain wall types ($\uparrow\downarrow$ and $\downarrow\uparrow$) move against the electron flow, we can conclude that SOTs are the dominating torques responsible for the DW displacement and that the DWs are all homo-chiral due to a finite DMI. 
\subsection{Controlling the domain wall velocity by an in-plane field}
\begin{figure}[htbp]
	\centering
	\includegraphics[width=80mm]{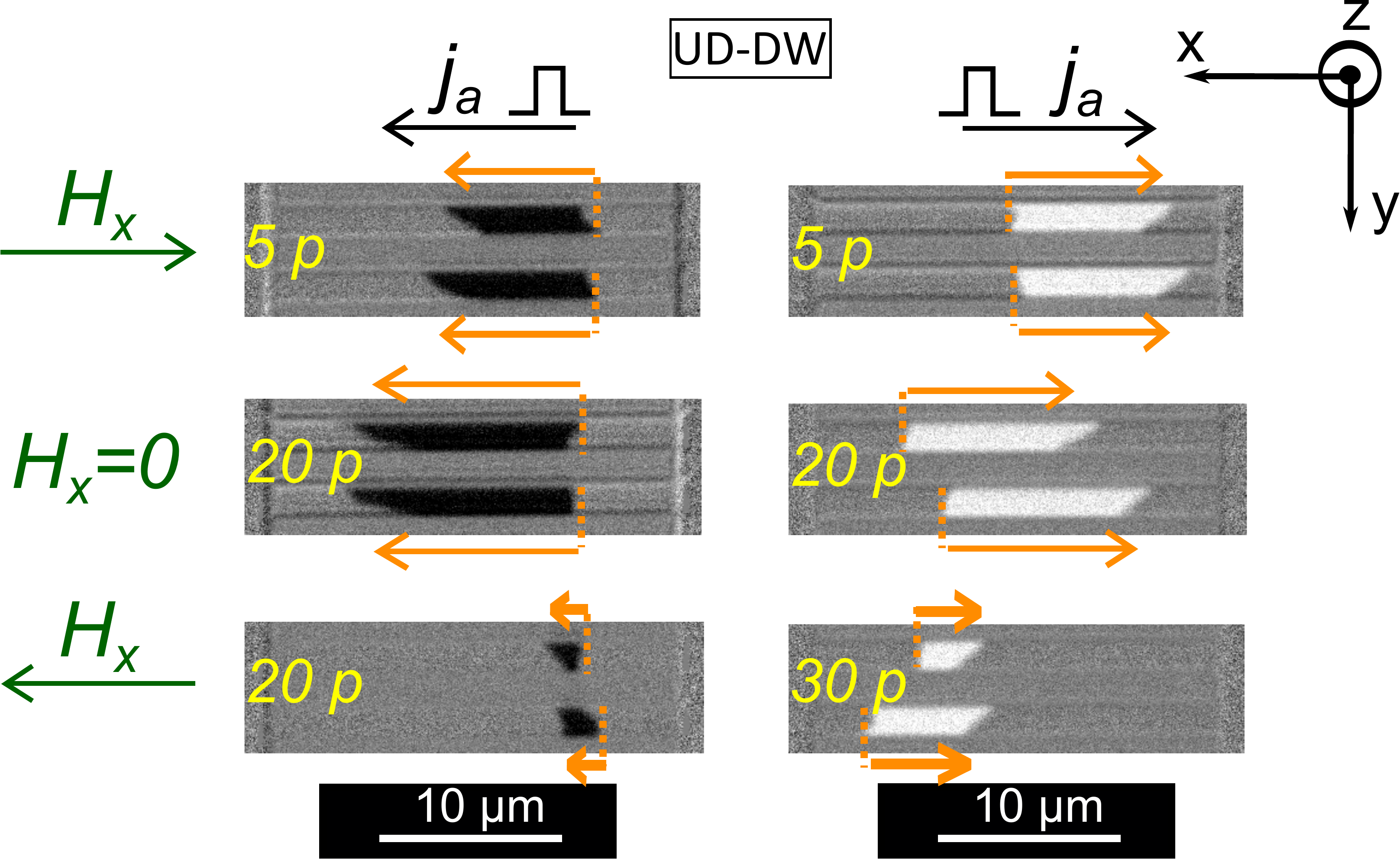}
	\caption{Current-induced motion of chiral DWs in the presence of a fixed in-plane longitudinal field, $\mathbf{H_{x}}$. The velocity of the DWs is observed to be reduced for ${H_{x}}>0$ and increased for ${H_{x}}<0$, for both ${j_{a}}>0$ and ${j_{a}}<0$. The gold dashed lines indicate the initial position of the DWs. The gold arrows indicate the DW direction of motion. The injected current density is $j_{a}=8.7\times10^{11}$~$A/m^{2}$, while the longitudinal field amplitudes are: $\mu_{0}H_{x}=-58$~mT, $\mu_{0}H_{x}=0$~mT, $\mu_{0}H_{x}=+58$~mT. The yellow numbers indicate how many pulses are used to generate the shown DW displacements. The different DW displacements in the two wires reflect the statistical spread of our observations. The images show current-induced motion of $\uparrow\downarrow$-DWs in the device with $t_{Co}=0.93$~nm.}
	\label{fig_vDW-Hx_protocol}
\end{figure}

\begin{figure*}[htbp]
	\centering
	  \subfigure[]
		  {\includegraphics[width=75mm]{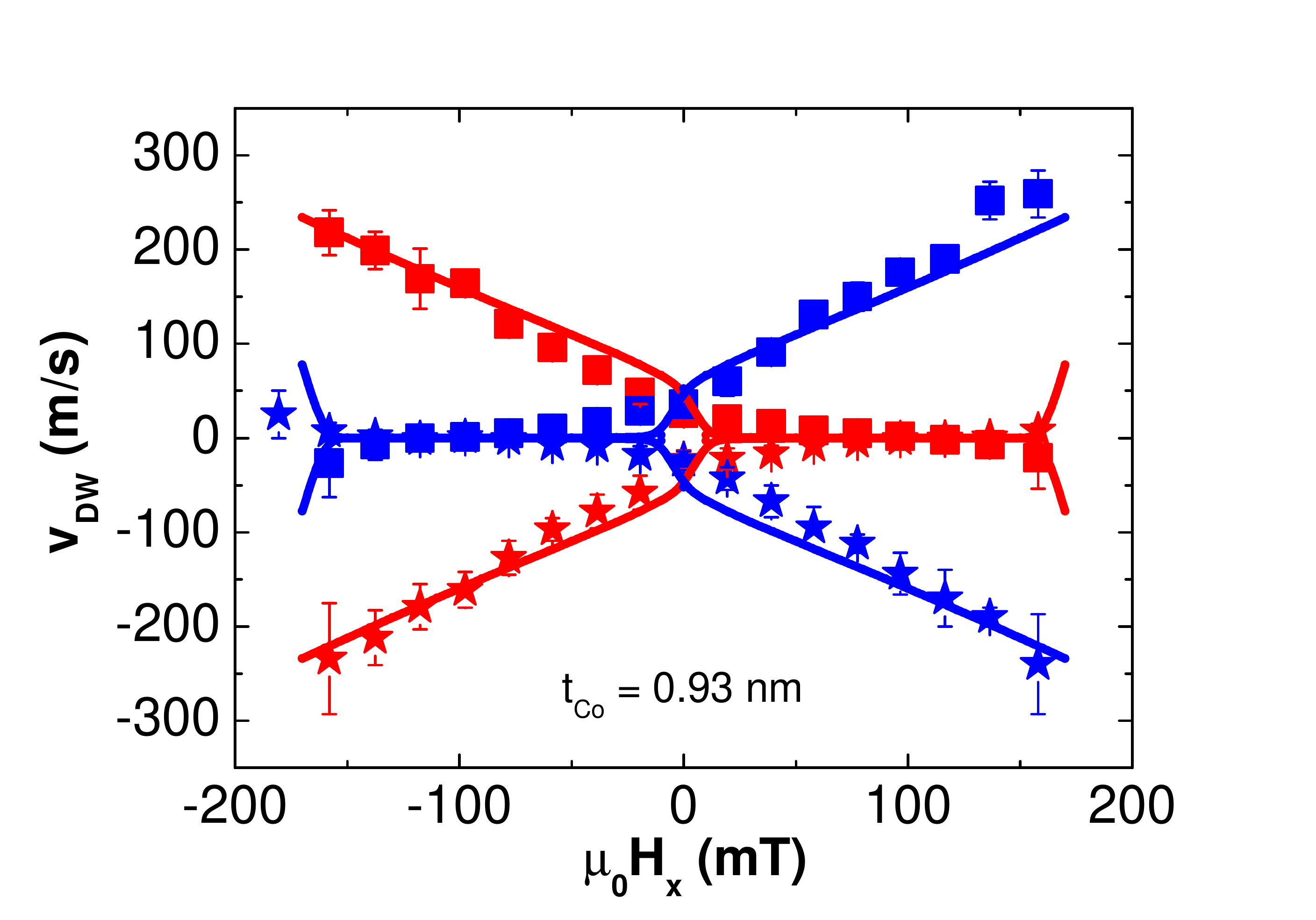}
		     \label{fig_vDW-Hx_1}}
		\subfigure[]
		  {\includegraphics[width=75mm]{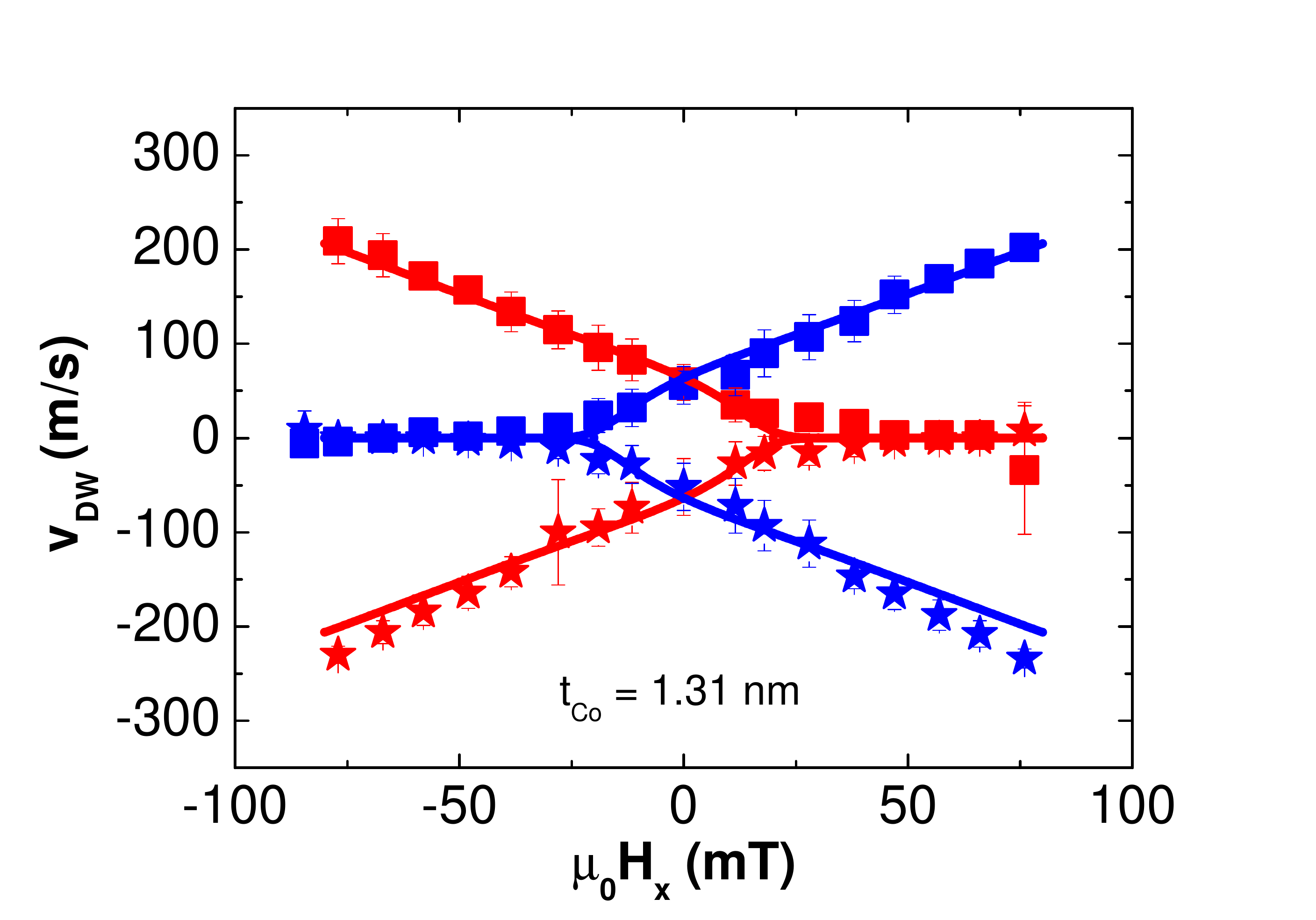}
		     \label{fig_vDW-Hx_3}}
		\subfigure[]
		  {\includegraphics[width=75mm]{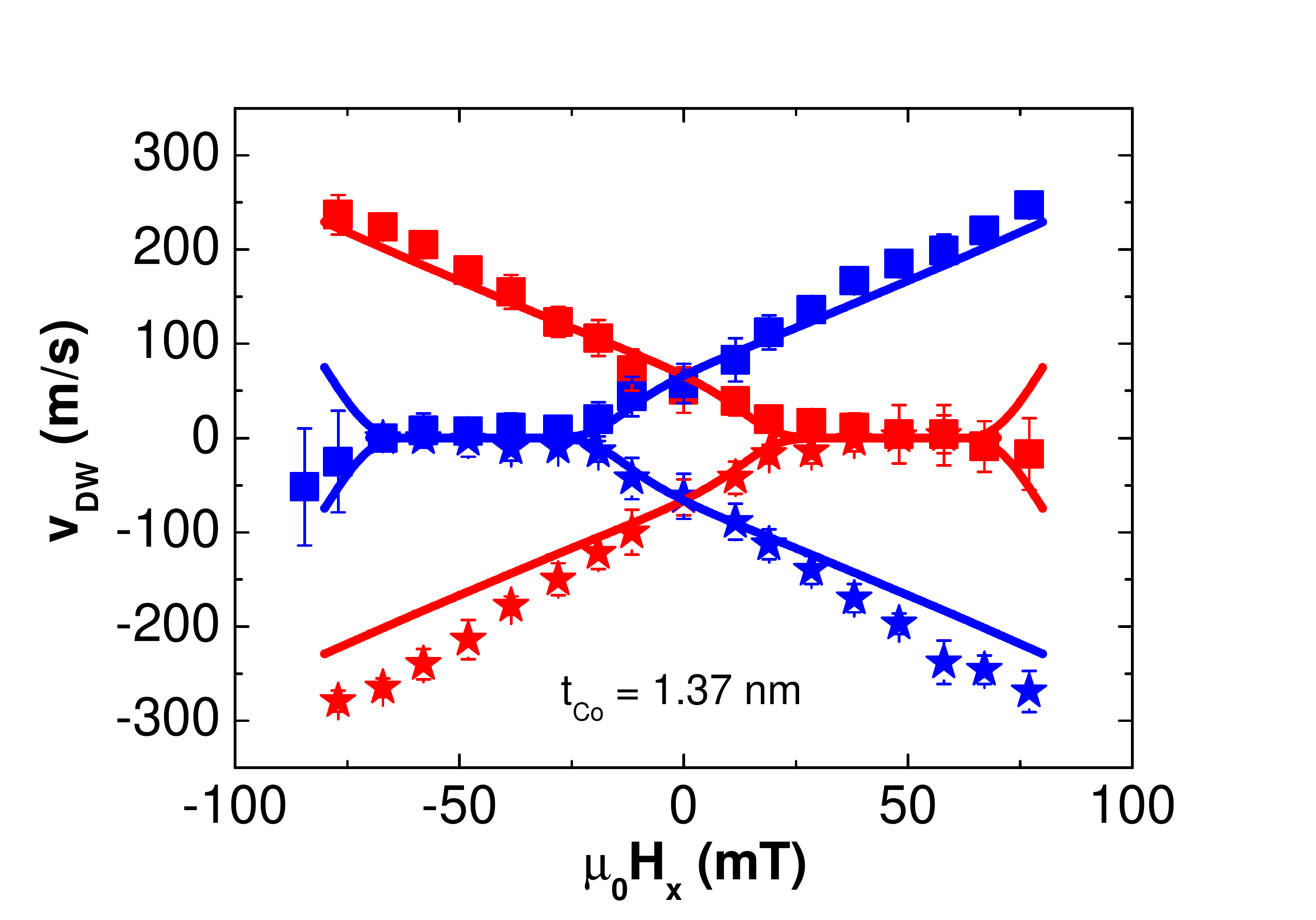}
		     \label{fig_vDW-Hx_4}}
		\subfigure[]
		  {\includegraphics[width=75mm]{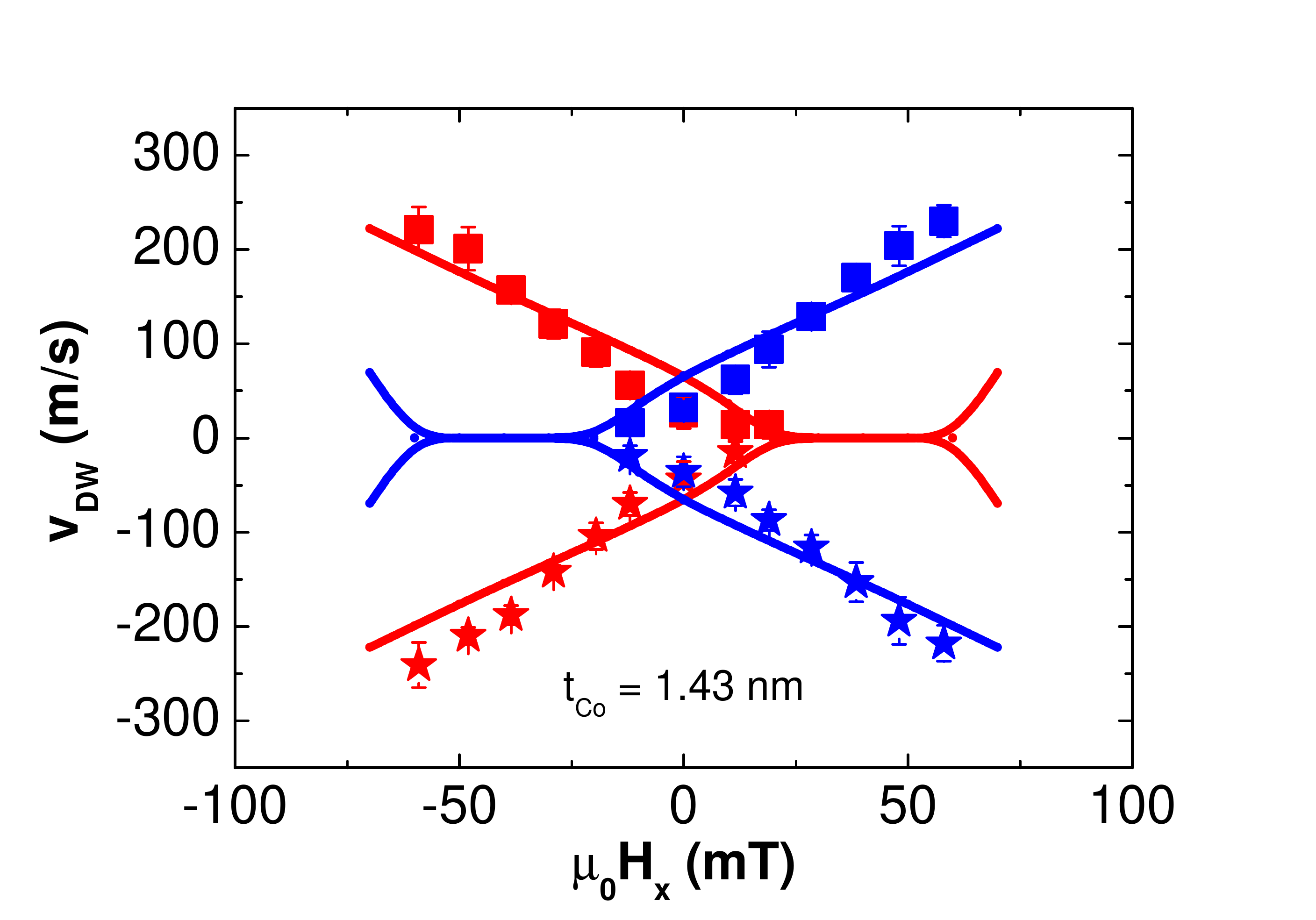}
		     \label{fig_vDW-Hx_5}}
	\caption{Symmetry of the velocity of chiral DWs. Average velocity, $v_{DW}$, of $\uparrow\downarrow$ (red symbols) and $\downarrow\uparrow$ (blue symbols) DWs as a function of $\mu_{0}H_{x}$, for the four different devices, for $j_{a}>0$ (squares) and $j_{a}<0$ (stars). The average velocities and the error bars (standard deviations) are calculated from several DW motion events for each magnetic field value. The solid lines are the fitting curves obtained by 1D model calculations showing a good agreement with the experimental data. (a) Data for $t_{Co}=0.93$~nm; $j_{a}=8.7\times10^{11}$~A/m$^{2}$. (b) $t_{Co}=1.31$~nm; $j_{a}=9.2\times10^{11}$~A/m$^{2}$. (c) $t_{Co}=1.37$~nm; $j_{a}=9.1\times10^{11}$~A/m$^{2}$. (d) $t_{Co}=1.43$~nm; $j_{a}=9\times10^{11}$~A/m$^{2}$.}
	\label{fig_vDW-Hx}
\end{figure*}

Next, we use CIDWM to determine the DMI in these samples. To this end, the DW velocity is measured as a function of an applied magnetic field along the wire axis ($\mathbf{H_{x}}$) for a fixed current density. The measurement protocol is the following. First, one type of DW ($\uparrow\downarrow$ or $\downarrow\uparrow$) is nucleated in each pre-saturated NW by current-induced magnetization switching \cite{miron2011perpendicular,loconte2014APL}. A typical switching pulse used in the experiment has a current density amplitude of $j_{a}\approx8\times10^{11}$~A/m$^{2}$ and a time duration of $\Delta{\tau_{p}}=40-50$~ns, assisted in the switching process by a fixed external longitudinal field of about 50~mT. Once the DWs are nucleated, they are displaced by the injection of a burst (n=1-50) of current pulses with a measured duration of 15~ns.\\The motion of $\uparrow\downarrow$-DWs in the presence of an applied longitudinal field is shown in Fig. \ref{fig_vDW-Hx_protocol}. The total DW displacement is strongly affected by the presence of a finite $H_{x}$. A positive field makes the $\uparrow\downarrow$-DWs move slower, while a negative field makes them move faster.
The measured average DW velocities as a function of the longitudinal field, $\mu_{0}H_{x}$, for all the devices are reported in Fig. \ref{fig_vDW-Hx} (symbols). Red (blue) symbols refer to $\uparrow\downarrow$- ($\downarrow\uparrow$-) DWs, while squares (stars) refer to ${j_{a}}>0$ (${j_{a}}<0$).\\As visible in Fig. \ref{fig_vDW-Hx_protocol} and Fig. \ref{fig_vDW-Hx}, while at zero-field the velocity of both types of DWs is the same, in the presence of a finite longitudinal field the two types of DWs move at different velocities. The change in the field amplitude affects differently the velocity of the two types of DWs, making it possible to obtain $\uparrow\downarrow$-DWs and $\downarrow\uparrow$-DWs moving in opposite directions, when the field amplitude is large enough. A symmetric behavior is observed for the velocity of the two DW types with respect to $H_{x}$, which can be described as: $v^{\uparrow\downarrow}_{DW}(j_{a},H_{x})=v^{\downarrow\uparrow}_{DW}(j_{a},-H_{x})$.
\\Considering an $\uparrow\downarrow$-DW, a sufficiently large positive $H_{x}$ slows them down, while a negative $H_{x}$ speeds the walls up. For very large positive $H_{x}$ the $\uparrow\downarrow$-DWs are also observed to change their direction of motion. However, differently from the case of the Ta$\backslash$CoFeB$\backslash$MgO system \cite{loconte2015PRB}, it is not possible in this case to access the regime of fast DW motion in the reversed propagation direction. Indeed, for in-plane fields larger than the ones shown in the graphs, local spontaneous magnetization reversal events \cite{torrejon2014interface,loconte2014APL} start to occur, making the DW motion measurement not possible anymore. The amplitude $H_{x}$ at which domains start to nucleate is observed to decrease with an increasing $t_{Co}$. This results in a reduction of the range of in-plane fields which can be used for the investigation of DW motion, moving from the thinnest to the thickest device.\\There is no single simple reason why the nucleation probability is observed to increase with increasing $t_{Co}$, but different effects can play a role. First, an increasing $t_{Co}$ results, for ultra-thin layers like the ones under investigation, in an increasing conductivity of the ferromagnetic layer \cite{li2000insitu}. Accordingly, the observed increasing nucleation probability could be linked with the increase of the current flowing in the ferromagnetic layer, due to an increasing Joule heating produced directly in the ferromagnetic material \cite{loconte2014APL}. Secondly, as reported in Fig. \ref{fig_characterization}, the magnetic anisotropy of the magnetic layer decreases as $t_{Co}$ increases. Thus, the required amplitude of the in-plane field, at a fixed current density, for driving the magnetization reversal in the NWs is expected to decrease with an increasing thickness \cite{lee2013threshold}. The nucleation process is observed to be particularly prevalent in the thickest investigated device, where already at applied fields of 10-20~mT reverse domains start to appear in the NWs after the current pulse injection. This results in the impossibility of observing any reversal in the DW motion direction, as shown in Fig. \ref{fig_vDW-Hx_5}.\\All that has been described above concerning $\uparrow\downarrow$-DWs is equally valid for $\downarrow\uparrow$-DWs for a symmetric reversal of the field $H_{x}$. In this case, a positive in-plane field speeds up DW motion, while a negative field slows them down (see Fig. \ref{fig_vDW-Hx}).\\These observations suggest strong spin-orbit torques acting in the materials stack, in combination with the presence of an interfacial Dzyaloshinskii-Moriya interaction \cite{emori2013current,ryu2014chiral,thiaville2012dynamics}. Having established that both SOTs and DMI are present, the key task is to determine the sign and the strength of DMI and SOTs as a function of the Co-layer thickness, as reported next.
\section{Thickness dependence of the Dzyaloshinskii-Moriya interaction}
\begin{figure}[htbp]
	\centering
	\includegraphics[width=80mm]{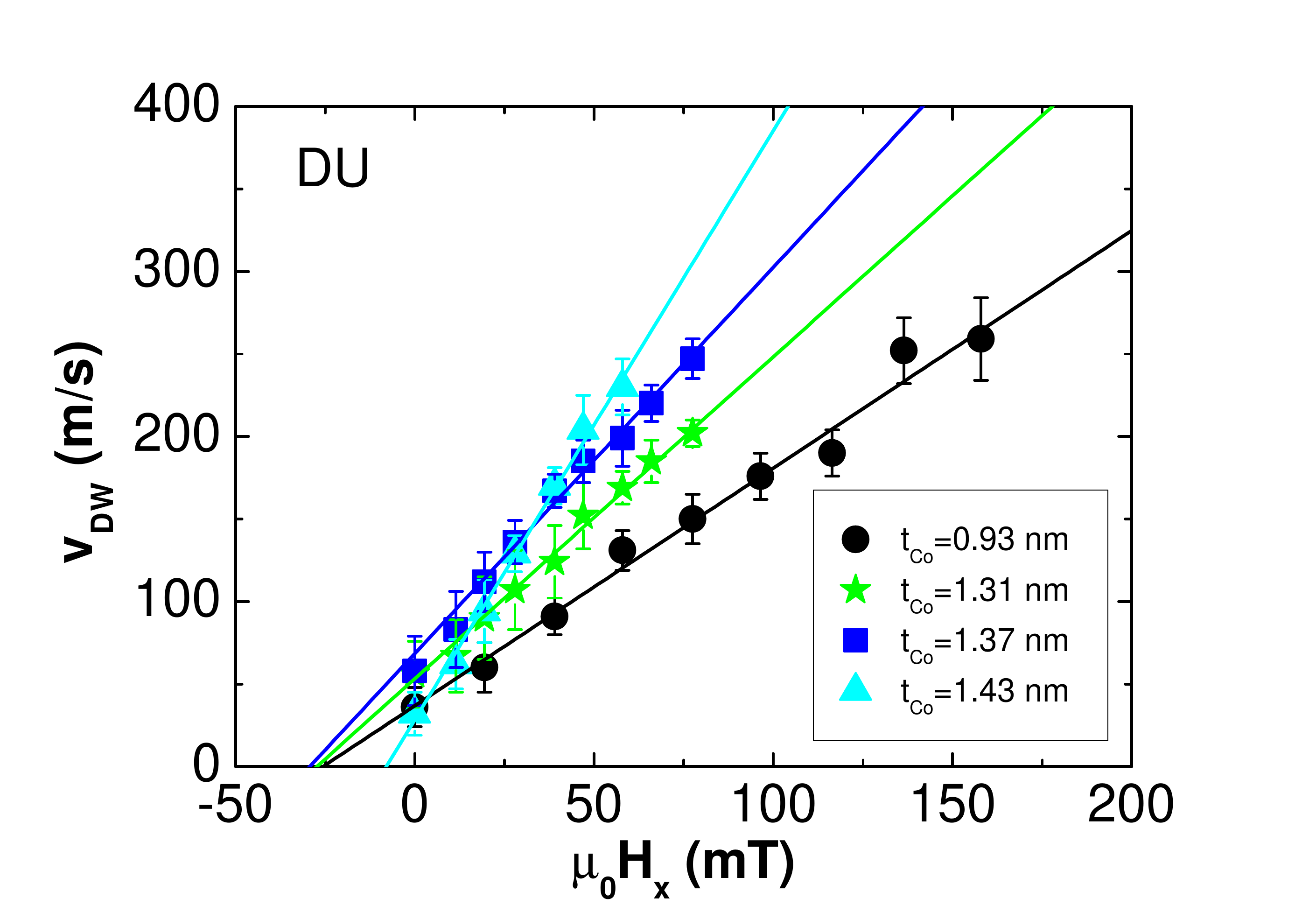}
	\caption{High velocity data points for the $\downarrow\uparrow$-DW with respect to $H_{x}$, for the four studied devices. The reported data points refer to measurements with $j_{a}>0$, while $H_{x}$ is parallel to the intrinsic N\'{e}el-component of the DW internal magnetization. The lines are the linear fitting curves for the experimental data. The slope is found to increase with increasing $t_{Co}$.}
	\label{fig_vDW-Hx_linFit}
\end{figure}

\begin{figure}[htbp]
	\centering
	  \subfigure[]
		  {\includegraphics[width=80mm]{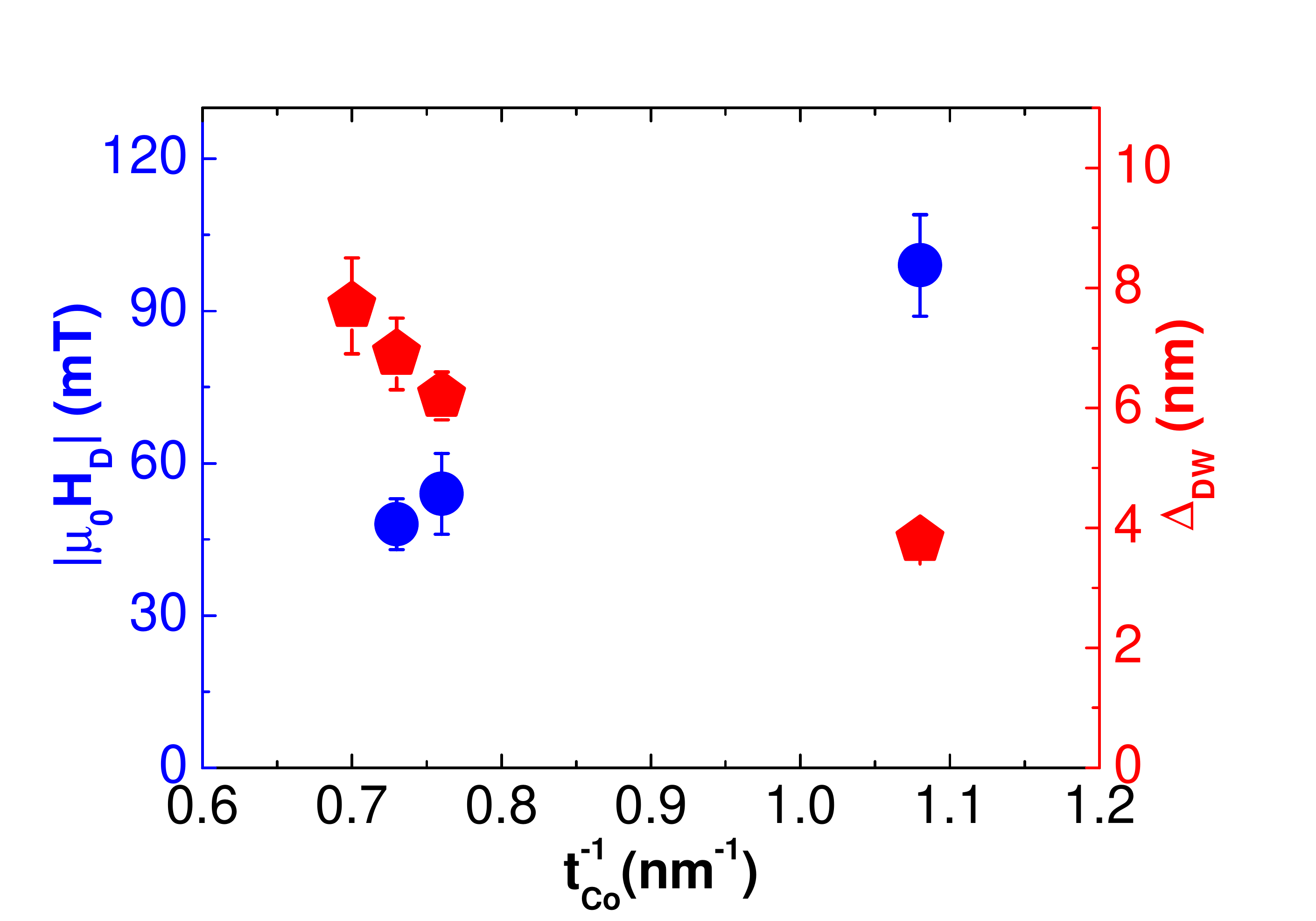}
		     \label{fig_DMI_1}}
		\subfigure[]
		  {\includegraphics[width=80mm]{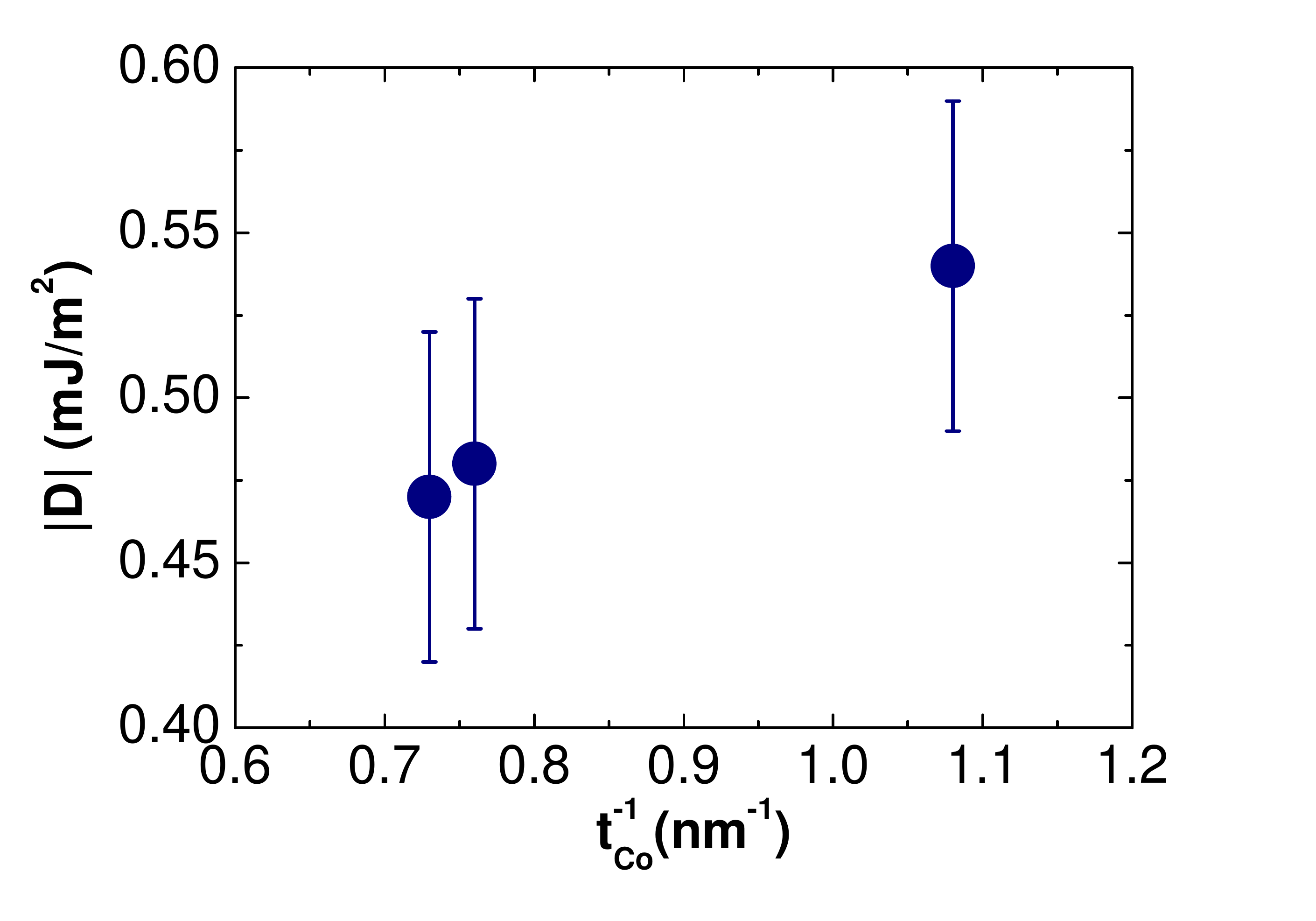}
		     \label{fig_DMI_2}}
	\caption{Extracted DMI fields and strengths. (a) Extracted effective DMI field, $|\mu_{0}H_{D}|$ (blue dots), and calculated DW parameter, $\Delta_{DW}$ (red pentagons), as a function of the inverse of the Co thickness. (b) Absolute value of the extracted DMI coefficients, $|D|$, as a function of the inverse of the Co thickness.}
	\label{fig_DMI}
\end{figure}

We use the CIDWM results to determine the DMI present in the different devices. In order to do so, the so-called stopping fields for the $\uparrow\downarrow$-DW and $\downarrow\uparrow$-DW need to be extracted. The stopping field is the external longitudinal field which needs to be applied in order to make the DW stop moving \cite{loconte2015PRB}. This can be extracted from the graphs shown in Fig. \ref{fig_vDW-Hx}.\\It is known that DMI stabilizes N\'{e}el DWs with the same chirality \cite{ryu2014chiral,thiaville2012dynamics,martinez2014current}. This can be interpreted as due to the presence of an effective DMI field, $\mathbf{H}_{D}$, along the longitudinal direction. Accordingly, the stopping field is the field needed to counteract the DMI field and turn the DW spin texture into the Bloch configuration, making the spin-orbit torques acting on the DW zero.\\The common approach used so far in order to extract the stopping fields from the CIDWM graphs is based on a linear fitting of the high velocity data points (for positive and negative $H_{x}$), for each $DW-j_{a}$ combination \cite{ryu2014chiral,loconte2015PRB}. The crossing point between the zero-velocity axis and the fitting curve would define the stopping field. However, it is not possible to use this approach in the present case. The reason is the lack of high velocity data points for one of the two field signs (positive, for $\uparrow\downarrow$-DWs; negative, for $\downarrow\uparrow$-DWs) in the reported graphs, as discussed above. A simple linear fitting of the high velocity data points (see Fig. \ref{fig_vDW-Hx_linFit}) results in the extraction of stopping field values that are outside the observed pinning ranges in Fig. \ref{fig_vDW-Hx}. According to that, an alternative approach to extract the stopping fields is used.
\\The range of in-plane magnetic fields where the DWs move at low velocities and become pinned is centered around the stopping field. Indeed, the damping-like SOT driving the DW motion is proportional to the N\'{e}el component of the DW, which in turn results from the interplay of the external field and the DMI field. So, the stopping field where the DW does not feel a force is equal in magnitude and opposite in sign compared to the DMI field \cite{emori2013current,ryu2014chiral,martinez2014current}. As a consequence, the pinning range can be expressed as: $[-H^{\uparrow\downarrow,\downarrow\uparrow}_{D}-\Delta{H_{x}}^{dep},-H^{\uparrow\downarrow,\downarrow\uparrow}_{D}+\Delta{H_{x}}^{dep}]$; where $H^{\uparrow\downarrow,\downarrow\uparrow}_{D}$ is the DMI effective field for $\uparrow\downarrow,\downarrow\uparrow$ DWs, $2\Delta{H_{x}}^{dep}$ the pinning range of longitudinal magnetic fields. Accordingly, the DMI effective field can be extracted as the center of the observed pinning ranges of $H_{x}$. Finally, the strength of the DMI can be obtained by the equation $D=\mu_{0}H_{D}M_{s}\Delta_{DW}$ \cite{thiaville2012dynamics,emori2014spin}.\\To determine the two stopping fields, ${H^{\uparrow\downarrow}_{x}}$ and ${H^{\downarrow\uparrow}_{x}}$, we first ascertain the magnetic field value, ${H}_{1}$, to which corresponds the highest measured $v_{DW}$ in the reversed direction. This velocity is used as the threshold velocity, $\tilde{v}_{DW}$, which defines the boundary between the low velocity regime (pinning regime) and the high velocity regime. Secondly, the data point with the same absolute value of the velocity, but with opposite sign is determined, together with the magnetic field at which it occurs, ${H}_{2}$. These two values of the applied magnetic field are used as extremes of the pinning range, and the corresponding stopping field, ${H}^{+,-}_{stop}$, is calculated as the arithmetic mean of the two. Two stopping fields are calculated for each type of DW (when this is possible), corresponding to $j_{a}>0$ (${H}^{+}_{stop}$) and $j_{a}<0$ (${H}^{-}_{stop}$), and the average stopping field for each DW type is obtained as ${{H}_{x}}^{\downarrow\uparrow,\uparrow\downarrow}=\frac{{H}^{+}_{stop}+{H}^{-}_{stop}}{2}$.\\The described protocol for the extraction of the effective DMI field is now applied to the experimental data reported in Fig. \ref{fig_vDW-Hx_1}, referring to the magnetic device with $t_{Co}=0.93$~nm. The extracted stopping field for the $\uparrow\downarrow$-DW is $\mu_{0}{H}^{\uparrow\downarrow}_{x}=99\pm10$~mT (error defined as $\frac{\mu_{0}{H}^{+}_{stop}-\mu_{0}{H}^{-}_{stop}}{2}$); while for the $\downarrow\uparrow$-DW $\mu_{0}{H}^{\downarrow\uparrow}_{x}=-99\pm9$~mT. Since ${{H}_{D}}^{\downarrow\uparrow,\uparrow\downarrow}=-{{H}_{x}}^{\downarrow\uparrow,\uparrow\downarrow}$, an effective DMI field of $\mu_{0}{H}_{D}=\frac{\mu_{0}{H}^{\uparrow\downarrow}_{D}-\mu_{0}{H}^{\downarrow\uparrow}_{D}}{2}=-99\pm10$~mT is obtained.\\In order to extract the effective DMI coefficient $D$, the DW width parameter needs to be calculated first. In the case of the thinnest device, $\Delta_{DW}=\sqrt{A/K_{eff}}=3.8\pm0.1$~nm (where the used value of the exchange stiffness for the Co layer $A=1.6\times10^{-11}$~J/m is chosen as reported in literature \cite{thiaville2012dynamics,hrabec2014measuring}). Accordingly, the effective DMI results to be $D=\mu_{0}{H}_{D}M_{s}\Delta_{DW}=-0.54\pm0.04$~mJ/m$^{2}$.\\The same process is repeated for $t_{Co}=1.31$~nm and $t_{Co}=1.37$~nm, extracting the respective effective DMI field and DMI strength, which are reported in Fig. \ref{fig_DMI_1} and Fig. \ref{fig_DMI_2}, respectively. For the device with thickness of $t_{Co}=1.43$~nm it is not possible to extract a reliable DMI value. The lack of observations of DW motion in the pinning regime, does not allow to employ the described DMI extraction method for that specific device. 
\section{Thickness dependence of current-induced spin-orbit torques}   
The injection of an in-plane current through a NM$\backslash$FM hetero-structure generates two different types of torques: a damping-like (DL) torque and a field-like (FL) torque, corresponding to two current-induced effective fields: $H_{DL}$ and $H_{FL}$, respectively. These torques are responsible for the efficient current-induced domain wall motion and magnetization switching observed in those hetero-structures \cite{miron2011perpendicular,miron2011fast,emori2013current,ryu2014chiral}. Here the thickness dependence of SOTs is obtained by two complementary approaches: on the one hand, the effective field moving the DWs is extracted by analyzing the CIDWM data by a collective-coordinate model; on the other hand, both the DL-field and the FL-field are characterized by second harmonic Hall measurements.
\subsection{Extraction of the domain wall motion effective field}
\begin{figure}[htbp]
	\centering
	\includegraphics[width=80mm]{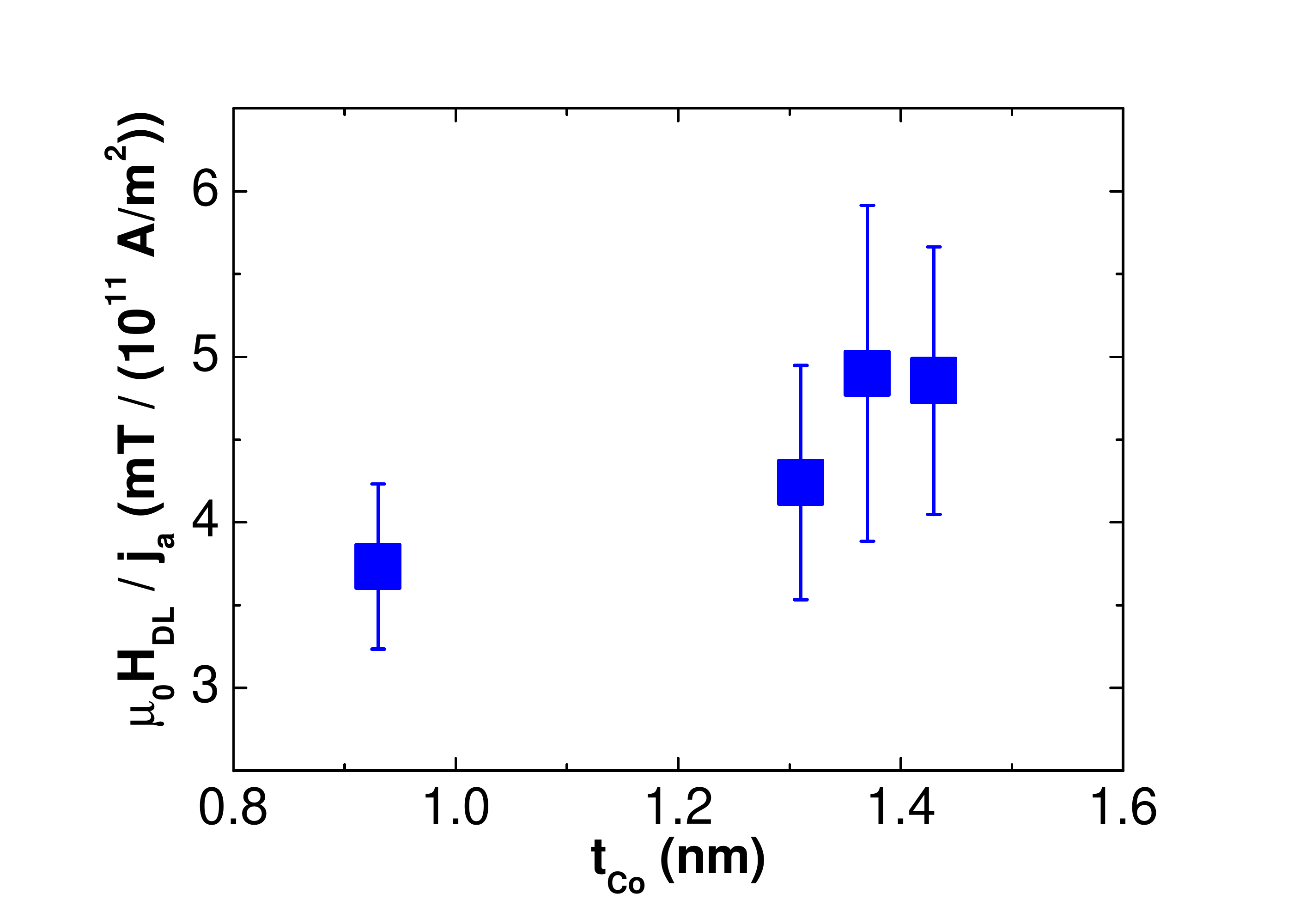}
	\caption{Domain wall motion effective field per current density, $H_{DL}/j_{a}$, as extracted by the 1DM calculations. The values of $H_{DL}/j_{a}$ shown here are the ones used for the generation of the fitting curves reported in Fig. \ref{fig_vDW-Hx}.}
	\label{fig_1DM_SOTs}
\end{figure}

To determine the spin-orbit torques acting on the DW, a collective-coordinate model (CCM) \cite{martinez2014current} based on the extension of the one-dimensional model (1DM) is employed to reproduce the experimental observations reported in Fig. \ref{fig_vDW-Hx} (see Appendix A for more details). In the framework of the CCM, the DW dynamics is described by three degrees of freedom: the position of the DW in the track, $q$, the in-plane angle of the DW magnetic moment with respect to the $x$-axis, $\phi$, and the angle defined by the normal to the DW surface with respect to the $x$-axis, $\chi$, describing the tilt of the DW plane.\\The action of the SOTs is equivalent to the presence of two effective magnetic fields: $H_{DL}$ and $H_{FL}$ \cite{emori2013current,ryu2014chiral,martinez2014current}. In the present work the FL-field is defined as $H_{FL}=\eta{H_{DL}}$, with $\eta$ being the proportionality factor between the two effective fields. However, in this section only the zero FL-field scenario ($H_{FL}=0$) is discussed, due to the fact that the DL-SOT is the main source of the domain wall motion. In particular, we find that the same final DW velocity is predicted by the CCM calculations with or without the inclusion of a finite FL-field into the calculations (more details can be found in Appendix B).\\The resulting CCM fitting curves are shown in Fig. \ref{fig_vDW-Hx} (solid lines). As it can be seen, they reproduce the experimental data well. The effective field ($H_{DL}$) is used as a free parameter in the calculations, while the DMI values used for the fitting procedure are the experimentally extracted ones. By using the best fitting curves, the amplitude and sign of the effective SOT-field is extracted, for each device. The corresponding current-field efficiency, $H_{DL}/j_{a}$, is reported in Fig. \ref{fig_1DM_SOTs} as a function of the Co thickness. The sign of the effective field is in agreement with a positive spin-Hall angle, if the SHE is assumed as the main source of the observed SOT. This is in agreement with what previously reported in literature for Pt-based systems \cite{emori2013current,ryu2014chiral}.\\As shown in Fig. \ref{fig_1DM_SOTs}, the extracted effective field is observed to increase for a Co thickness between 0.9 and 1.4~nm and then to level off around $H_{DL}/j_{a}=5$~[mT/($10^{11}$A/m$^{2}$)] for larger thicknesses. This indicates that the effective SOT-field generating the observed DW motion is initially scaling with the ferromagnetic layer thickness and then becomes independent of it.     
\subsection{Extraction of effective SOT-fields by second harmonic technique} 
\begin{figure*}[htbp]
	\centering
	  \subfigure[]
		  {\includegraphics[width=75mm]{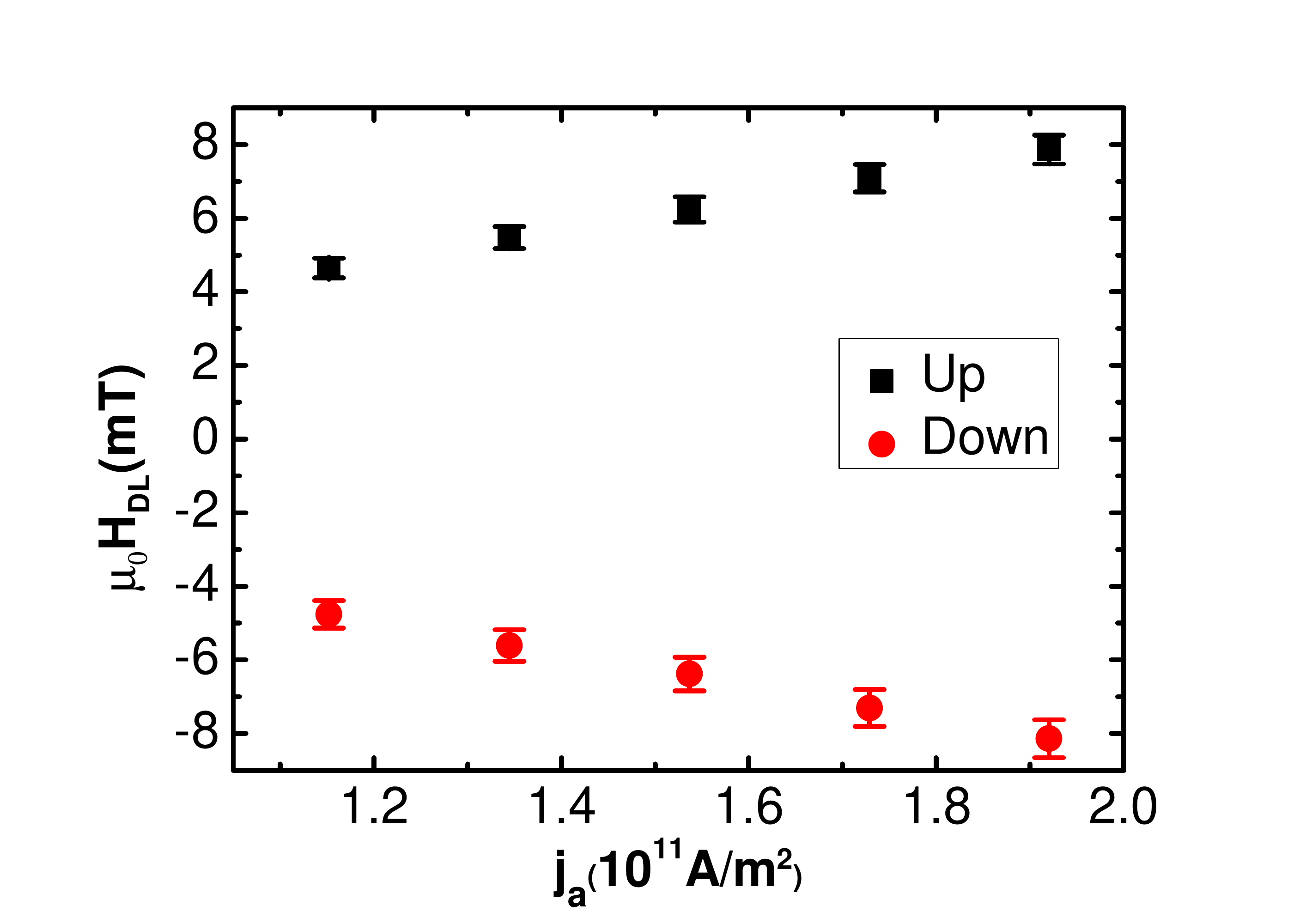}
		     \label{fig_SL-SOT}}
		\subfigure[]
		  {\includegraphics[width=75mm]{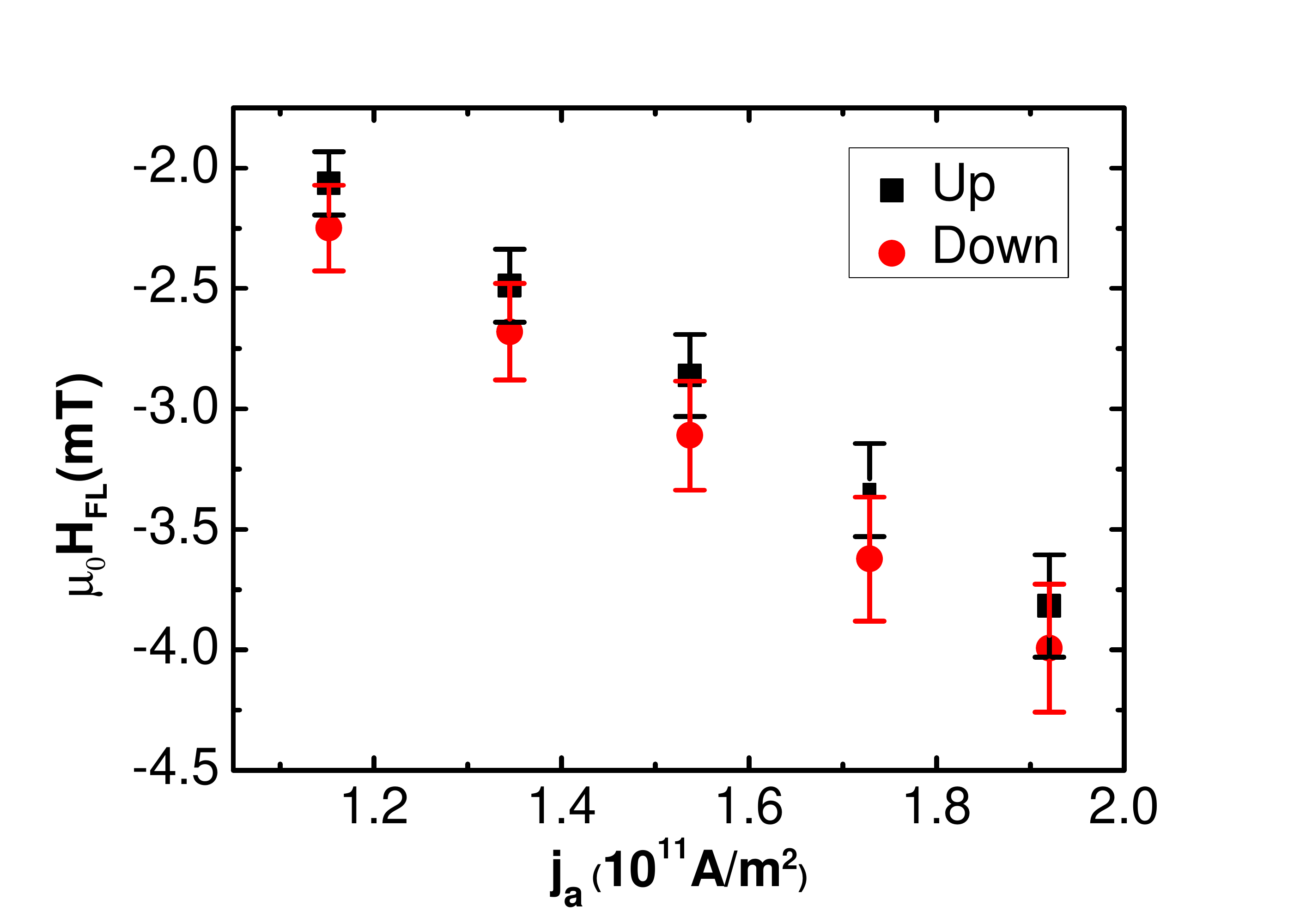}
		     \label{fig_FL-SOT}}
	\caption{SOT effective fields as a function of the current density, for the device with $t_{Co}=1.37$~nm. (a) Longitudinal effective field (DL-field). (b) Transverse effective field (FL-field). The calculated Oersted field due to the current flowing into the Pt layer is around $H_{Oe}/j_{a}=0.3$~[mT/($10^{11}$A/m$^{2}$)], an order of magnitude smaller than the effective FL-field reported here. Black squares (red dots) correspond to magnetization Up (Down).}
	\label{fig_SOTs}
\end{figure*}

\begin{figure}[htbp]
	\centering
	\includegraphics[width=80mm]{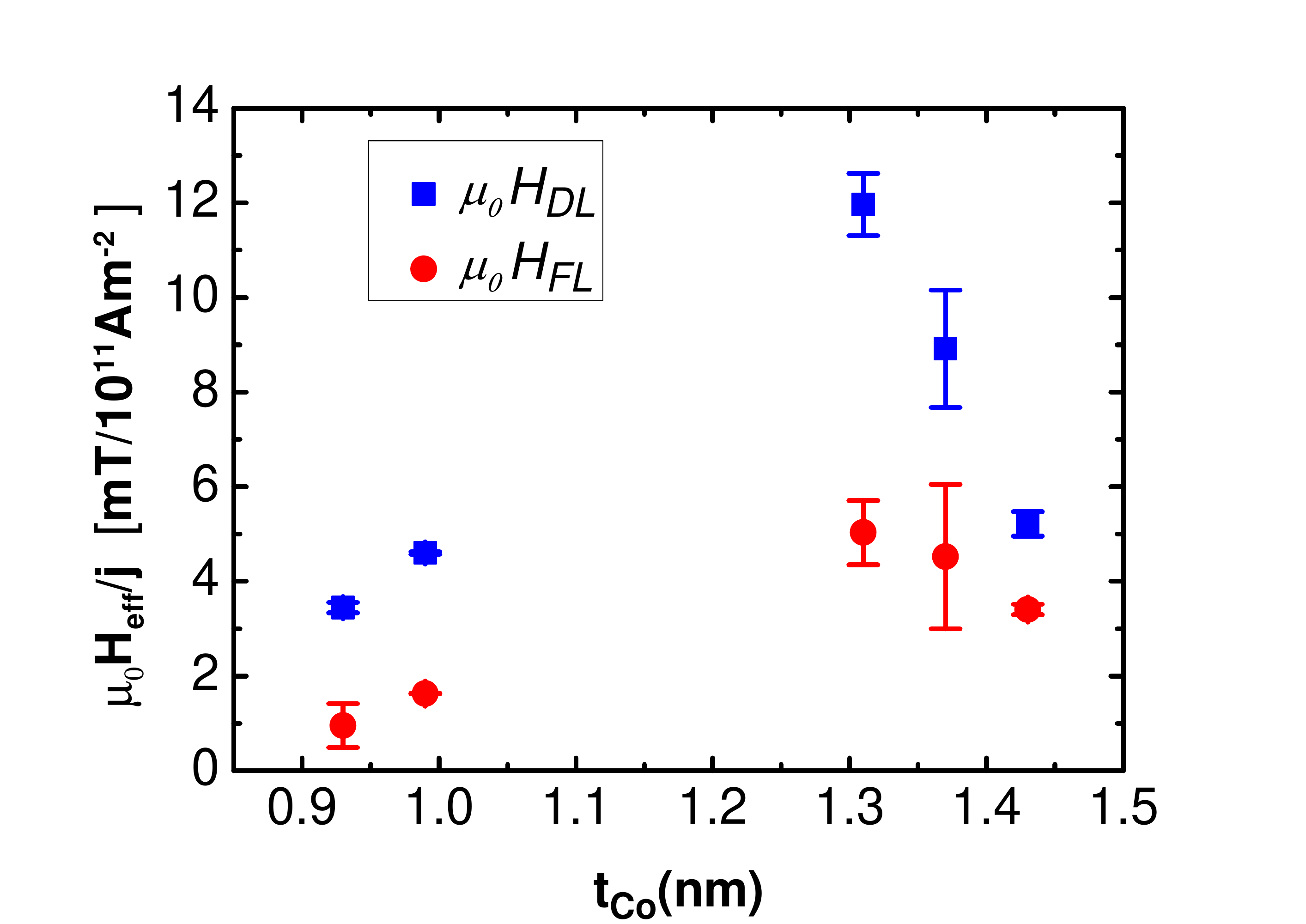}
\caption{Longitudinal ($H_{DL}$, black squares) and transverse ($H_{FL}$, red dots) effective fields as a function of the ferromagnetic thickness.}
\label{fig_SOT-vs-tCo}
\end{figure}

Next, we use the established $2\omega$ technique \cite{garello2013symmetry,pi2010tilting,hayashi2014quantitative,lee2014spin} to study the current-induced SOTs in Pt$\backslash$Co$\backslash$AlO$_{x}$ Hall cross devices. In Fig. \ref{fig_SOTs} the measured current-induced effective fields along the longitudinal (Fig. \ref{fig_SL-SOT}) and transverse (Fig. \ref{fig_FL-SOT}) direction are shown with respect to the injected current, for one of the studied devices. Both effective fields are found to scale linearly with the current amplitude.\\In order to learn more about the microscopic origins of SOTs in the present materials system, their ferromagnetic thickness dependence is studied. Accordingly, the current-induced SOTs are measured for different devices with a Co thickness of: 0.93~nm, 0.99~nm, 1.31~nm 1.37~nm and 1.43~nm.
\\The measured effective fields as a function of the Co thickness are reported in Fig. \ref{fig_SOT-vs-tCo}. First of all, the two effective fields are found to have the same qualitative dependence on the Co thickness, suggesting a possible common origin for such a dependence. Second, the two effective fields are observed to first increase and then decrease with the Co thickness, clearly indicating a more complicated character than a simple interface-like one. Indeed, in the pure interface-like case we would observe a simple 1/t dependence, which is clearly not the case here. Third, the DL-field is found to be always larger than the FL-field, for all the investigated Co thicknesses. In general, in [heavy metal]$\backslash$ferromagnet systems the primary origin of the DL-torque is usually attributed to the SHE due to the large SOC characterizing the heavy metal \cite{ryu2014chiral,liu2012current}. Accordingly, the spin-Hall effect is most probably an important source of both SOTs measured here, where the weaker FL-field is generated by the precession, around the exchange field, of the itinerant spins diffusing in the ferromagnetic layer \cite{haney2013current}. This interpretation is also in agreement with the observed non-monotonous thickness dependence. The SHE-induced spin-current diffuses in the ferromagnet and interacts with the local magnetization by generating the two SOTs. The length scale defining the thickness dependence of the corresponding effective fields is the transverse spin diffusion length in Co \cite{shpiro2003self}, reported to be around 1.2 nm \cite{ghosh2012penetration}. Indeed, after diffusing across the ferromagnet for a distance equal to the transverse spin-diffusion length, the spin-current is absorbed and no further effect on the magnetization beyond this thickness is produced. Accordingly, in the first spin diffusion length the effective fields build up, then decay with a further increase of the Co thickness. While this qualitatively describes the measurements, we cannot rule out further origins that would be influenced by an identical effect, resulting in the similar thickness dependence reported here.    
\section{Discussion} 
\subsection{Extracted DMI}
\begin{figure}[htbp]
	\centering
	\includegraphics[width=80mm]{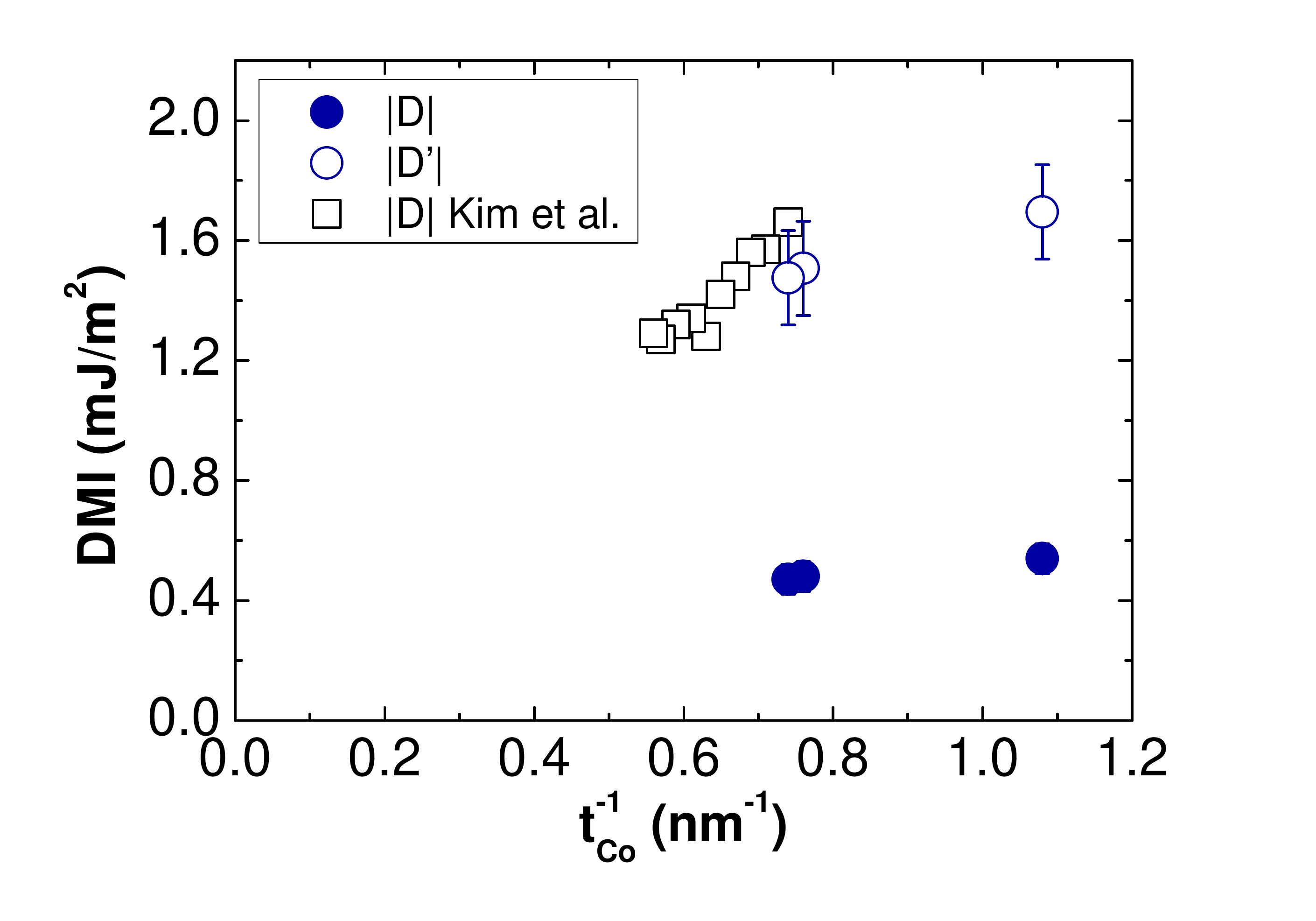}
	\caption{Comparing the extracted DMI strength with the results from BLS measurements. The absolute value of the extracted DMI coefficients, $|D|$ (solid dots), is compared with the values obtained by BLS by Kim et al. \cite{kim2015improvement} (open squares). The latter result to be about 3 times larger then the former. The open dots are the values of an arbitrarily defined $|D^{'}|=\pi|D|$.}
	\label{fig_DMI_total}
\end{figure}

In multilayer systems like the one discussed here, the DMI is predicted to originate from the interface between the heavy metal (Pt) and the ferromagnetic material (Co) \cite{yang2015anatomy}. Accordingly, the DMI is expected to be an interface-like effect, where its effective strength scales with the inverse of the ferromagnetic thickness. In Fig. \ref{fig_DMI_1} the measured DMI fields (blue spheres) are shown to be proportional to $t^{-1}_{Co}$. Furthermore, the extracted values of ${H}_{D}$ are in agreement with what has already been reported in literature for CIDWM experiments \cite{ryu2014chiral,safeer2016spin}. Safeer et al. \cite{safeer2016spin} reported a 100~mT stopping field for DW motion parallel to the current flow in a Pt(3~nm)$\backslash$Co(0.6~nm)$\backslash$AlO$_{x}$(2~nm) sample. Ryu et al. \cite{ryu2014chiral} reported a 140~mT stopping field in Pt(1.5~nm)$\backslash$[Co(0.3~nm)$\backslash$Ni(0.7~nm)$\backslash$Co(0.15~nm)] NWs. Accordingly, the range of stopping fields reported here is in line with previously reported values for other material systems with a Pt buffer layer, even though in those previous reports no systematic thickness dependence, as we provide here, is given.\\In Fig. \ref{fig_DMI_1} we also show the calculated DW width parameter, $\Delta_{DW}=\sqrt{A/K_{eff}}$ (red pentagons), as a function of $t^{-1}_{Co}$. Using the extracted ${H}_{D}$ and the calculated $\Delta_{DW}$ for calculating the DMI strength, results in a DMI that scales with the inverse ferromagnetic thickness (see Fig. \ref{fig_DMI_2}). This is in agreement with an interfacial DMI scenario, where the effective DMI strength is expected to be proportional to $1/t_{FM}$: $D\propto{\frac{D_{int}}{t_{FM}}}$, with $D_{int}$ being the pure interfacial DMI. However, a linear fitting of the data points (not shown here) does not generate a linear curve that crosses the origin of the axes. The crossing for $t^{-1}_{Co}\longrightarrow0$ would happen at a finite value of the ordinate axis. These observations are in agreement with what reported in Ref. \cite{cho2015thickness} for a Pt$\backslash$Co system, and in Ref. \cite{nembach2015linear} for a Pt$\backslash$Ni$_{80}$Fe$_{20}$ system. Cho and co-authors \cite{cho2015thickness} observed a linear dependence of the effective DMI strength on $t^{-1}_{Co}$ for $t^{-1}_{Co}>0.5$~nm$^{-1}$ (similar to the present study), and a rapid drop of it for $t^{-1}_{Co}<0.5$~nm$^{-1}$, going towards zero at $t^{-1}_{Co}\longrightarrow0$. Nembach and co-authors \cite{nembach2015linear} obtained similar results and attributed their observations to a non-trivial dependence of the interfacial DMI strength, $D_{int}$, on $t^{-1}_{FM}$. In the case reported here, devices with thicknesses larger than the studied ones exhibit a weak PMA and so the formation of a multi-domain state. This prevents the study of CIDWM and so the extraction of DMI, so that here we do not probe the expected faster decay of the effective DMI with respect to the inverse ferromagnetic thickness in thicker devices.
\\Moving the analysis from a qualitative to a quantitative level, we compare now the DMI strengths extracted here with the values obtained by Kim and co-authors \cite{kim2015improvement}, who carried out BLS measurements on the very same material stack used for the patterning of the devices reported in this manuscript. In Fig. \ref{fig_DMI_total} the DMI values by Kim et al. \cite{kim2015improvement} (squares) and our values (dots) are reported in the same graph. From the graph it is clearly visible that the values of $|D|$ (solid dots) obtained by CIDWM are in quantitative disagreement with the ones obtained by BLS. The two differently extracted values are off by about a factor 3. Indeed, if we multiply the extracted DMI values by a factor $\pi$ (arbitrary choice, any value between 3 and 4 could have been chosen), the values of $|D^{'}|=\pi|D|$ (open dots) are very close to the BLS values.  
This could indicate that the procedure of comparing the values of DMI extracted by different techniques needs to be adjusted. If, on the one hand, the $|D|$ obtained here are in good agreement with the outcome of other CIDWM-based measurements \cite{ryu2014chiral,safeer2016spin}, on the other hand $|D^{'}|$ is in good agreement with the values obtained by BLS measurements on the same sample \cite{kim2015improvement}. These results clearly motivate a theoretical analysis about the comparability of DMI measurements obtained with different experimental techniques. Similar differences were also found by Soucaille et al. \cite{soucaille2016probing}, where the DMI values extracted by DW creep motion and BLS studies are compared, for several materials stacks with low Gilbert damping ($\alpha=0.015-0.039$).\\One possible interpretation proposed by Soucaille et al. for the different values of DMI extracted by the two experimental techniques is the Gilbert damping experienced by the DW during the creep motion. They observed a larger discrepancy between the two techniques' results for materials stacks with a smaller damping factor, suggesting a better agreement for systems with large damping. However, despite the very large damping factor usually reported for Pt$\backslash$Co$\backslash$AlO$_{x}$ ($\alpha=0.1-0.3$ \cite{belmeguenai2015interfacial,schellekens2013determining}) we do observe a quantitative discrepancy between the DMI values extracted by CIDWM and the ones obtained by BLS. This rules out the small damping as the key factor generating the discrepancy between the CIDWM- and the BLS-extracted DMI values.\\A second possible interpretation of such a discrepancy relies on the different type of physical process probed by the two techniques. In the BLS measurements spin waves with wavelengths of hundreds of nanometers are probed on an area of tens of microns (defined by the laser spot) \cite{belmeguenai2015interfacial,kim2015improvement}. On the other hand, in CIDWM experiments the object at the center of the study is the DW, whose width is usually $<10$~nm in PMA systems. Accordingly, the DW motion is much more sensitive to the local variations in magnetic properties such as anisotropy, magnetization and DMI due to grain boundaries, interface roughness etc., while the spin waves probed in the BLS measurements are affected by the average values of those physical properties. Indeed, it has been suggested \cite{soucaille2016probing} that pinning sites could be characterized by a weaker DMI compared to the defect-free regions of the sample, which might help to explain the lower DMI probed by the DW compared to thermal magnons.\\A final possible interpretation for the lower DMI extracted by the CIDWM study relies on the presence of a non-negligible STT, even though for a related system no significant STT was claimed \cite{emori2013current}. It is known that the standard STT drives the DW motion along with the electron flow, unless exotic materials properties like a negative non-adiabaticity parameter characterize the system under study. However, to the best of our knowledge, no experimental observations of a negative non-adiabaticity parameter, in Pt$\backslash$Co$\backslash$AlO$_{x}$ and related systems, has ever been reported so far. Accordingly, the presence of a finite STT would result in the observation of a stopping field that is smaller than the actual DMI field (as previously reported by Torrejon et al. \cite{torrejon2014interface}). Nevertheless, if this was the main origin of the observed discrepancy, the strength of the STT needed to generate such effect would be $\approx\frac{2}{3}\approx0.7$ times the actual SOT's strength. This would correspond to a $70\%$ underestimation of the actual strength of the SOT by the used 1D model calculations. In the scenario of a simple model based on the SHE-induced SOT, where $H_{DL}=\frac{\hbar\theta_{SH}j}{2e\mu_{0}M_{s}t_{FM}}$, this would result in the extraction of a max SOT efficiency of $\theta^{max}_{SH}\approx0.9-1$. Such a large SOT efficiency has never been reported before in the study of CIDWM, and thus, most likely, this cannot explain our results. So the reported discrepancy between CIDWM- and BLS-extracted DMI values might be understood if a combination of all the three interpretations reported above is taken into account. However, it also shows that care needs to be taken when comparing values of the DMI determined by different techniques. Furthermore, this should encourage future comparative studies of different systems using different approaches and in particular theoretically analyze the different probed properties to identify the origin of the discrepancies, which will be reserved for a future study.
\subsection{Extracted SOTs}
When comparing the effective SOT-fields ($H_{DL}$) extracted by the CCM on the one side and the $2\omega$ technique on the other side, two major observations can be made. First, the effective fields obtained by the two techniques are of comparable magnitude. Second, while the effective fields extracted by the DW motion study are all contained in a range of values as large as the 25\% of the largest extracted value (see Fig. \ref{fig_1DM_SOTs}), for the effective fields obtained by Hall measurements the largest value is 4 times as large as the smallest one (see black squares in Fig. \ref{fig_SOT-vs-tCo}). Furthermore, while the effective field increase with increasing thickness is seen by both techniques, the second harmonics technique detects also a stronger decrease for large thicknesses that is not directly visible in the DW motion results.\\There may be several possible explanations for the different thickness dependence of the DL-field extracted by the two experiments, and here we discuss two of them. The first explanation is based on the symmetry of the SOTs. It is well known that SOTs depend on the polar angle (angle between the magnetic moment and the z-axis) \cite{garello2013symmetry,qiu2014angular}, and the thickness dependence of the same SOT (DL-torque in this case) does not need to be necessarily the same for different polar angles. Furthermore, the magnetic moment in a DW is characterized by a polar angle of 90 deg (contained in the x-y plane of the sample), while the magnetization in the Hall measurements is almost collinear to the z-axis ($\uparrow$ or $\downarrow$ state). Accordingly, the observed different results obtained by CIDWM and $2\omega$ measurements can be interpreted as a polar angle-dependent scaling of the DL-field with the ferromagnetic thickness. A second possible interpretation is based on the different spin structures that are probed in the two experiments. In the DW motion study, the effective SOT-field acts on a highly non-collinear spin texture, while in the Hall measurements the probed magnetic state is a single domain. Accordingly, the spin-transfer process in the two cases can be different, resulting in the extraction of different effective field amplitudes. Furthermore, the DMI present in this system plays a key role in the DW dynamics \cite{yang2015domain}, while it does not affect the current-induced dynamics of the saturated magnetic state. Finally, the 1D-model used to analyze the torques exerted in the domain wall motion is limited by the virtue that it does not capture the evolution of the structure of the domain walls during their motion and this could lead to a limited understanding of the precise magnitude of the torques. In a future work, using independent determination of the damping and more refined modeling might lead to a more quantitative agreement.\\In conclusion, we observe that the analysis of the results  of $2\omega$ and DW motion experiments provide us with effective SOT-fields that are of similar magnitude. However a detailed correlation between the trend of the effective fields measured by the two methods calls for further investigation, which is beyond the scope of this work.  
\section{Conclusions}
A detailed characterization of the thickness dependence of Dzyaloshinskii-Moriya interaction (DMI) and spin-orbit torques (SOTs) in Pt$\backslash$Co(t)$\backslash$AlO$_{x}$ was obtained by mean of two different experimental techniques: current-induced domain wall motion and second-harmonic Hall measurements. The sign and strength of the DMI are extracted by measuring the domain wall motion stopping fields. A negative DMI, corresponding to the presence of left-handed homo-chiral DWs, is observed to decrease in strength with an increasing Co thickness. This confirms that the DMI originates from the Pt$\backslash$Co interface and its measured effective strength decreases for thicker ferromagnetic layers. The extracted DMI strengths are in agreement with values reported in previous current-induced DW motion studies on similar materials stacks, however they are quantitatively different from the values extracted for the very same sample by Brillouin light scattering (BLS) measurements. The DMI values presented here are about a factor 3 smaller than the values extracted by BLS. This quantitative disagreement highlights that care has to be taken when comparing the results of the two techniques and the differences possibly originate from the different length scales and physical processes probed in the two experiments.\\From the analysis of the DW motion using a collective-coordinates model we extracted the driving SOT-field, while the symmetry and magnitude of both damping-like and field-like SOT-fields acting on a magnetic single domain state were extracted by $2\omega$ Hall measurements. The effective damping-like SOT-field driving the DW motion is observed to increase with an increasing ferromagnetic layer thickness up to $t_{Co}\approx1.4$~nm. From the $2\omega$ measurements both effective fields acting on the homogeneous spin structure are found to initially increase up to a Co thickness of about 1.3 nm, and then to decrease with a further increase of the ferromagnetic thickness. The differences in the thickness dependence of the effective fields obtained by the two techniques can have different origins: the non-ability of the 1D-model to capture the real dynamics of the DW internal magnetization during its motion, or the different parts of the polar angle dependence of the torques probed in the two experiments.\\The similar non-monotonic thickness dependence observed for the damping-like and field-like effective fields suggests a possible common origin for the two fields: the spin-Hall effect in the Pt bottom layer. In this scenario, the final effective fields are defined by the transverse spin diffusion length in Co, which is expected to be around 1.2 nm. Accordingly, for Co thicknesses smaller than the transverse spin diffusion length the effective fields are observed to increase, while for Co thicknesses larger than the spin diffusion length the effective fields are observed to decrease. However, while qualitatively the behavior fits a dominating SHE origin we cannot rule out additional effects that affect the two torques similarly leading to a similar thickness dependence. Finally, the qualitatively different thickness dependence of DMI and SOTs shows that both effects clearly do not have a common origin in our investigated system.   
\section*{Acknowledgments}
We acknowledge support by the Graduate School of Excellence Materials Science in Mainz (MAINZ) GSC 266, Staudinger Weg 9, 55128, Germany; the DFG (KL1811, SFB TRR 173 Spin+X); the EU (IFOX, NMP3-LA-2012 246102; MASPIC, ERC-2007-StG 208162; MultiRev ERC-2014-PoC 665672; WALL, FP7-PEOPLE-2013-ITN 608031) and the Research Center of Innovative and Emerging Materials at Johannes Gutenberg University (CINEMA). E. M. acknowledges the support by project MAT2011-28532-C03-01 from Spanish government and project SA163A12 from Junta de Castilla y Leon. C.-Y. You acknowledges the support by NRF-DFG Collaborative Research Program (No. 2014K2A5A6064900). J.-S. Kim acknowledges the support by the Leading Foreign Research Institute Recruitment Program (No. 2012K1A4A3053565) through the National Research Foundation of Korea.
\section*{Appendix A: Collective-coordinates model for CIDWM}
In the framework of the collective-coordinates model (CCM) the DW dynamics is described by three degrees of freedom: the position of the DW in the track, $q$, the in-plane angle of the DW magnetic moment with respect to the $x$-axis, $\phi$, and the angle defined by the normal to the DW surface with respect to the $x$-axis, $\chi$, describing the tilt of the DW surface. The system of equations governing the DW dynamics reads:
\begin{equation} \label{1DM_1}
\begin{split}
  (1+\alpha^{2})\frac{cos\chi}{\Delta}\frac{dq}{dt} & =\frac{Q\gamma_{0}}{2}[-H_{K}sin2(\phi-\chi)-Q{\pi}H_{D}sin(\phi-\chi)+{\pi}H_{x}sin\phi-{\pi}H_{FL}cos\phi] \\ 
	& +\alpha\gamma_{0}[H_{pin}(q)+Q\frac{\pi}{2}H_{DL}cos\phi],
\end{split}
\end{equation}
\begin{equation} \label{1DM_2}
\begin{split}
  (1+\alpha^{2})\frac{d\phi}{dt} & =-\frac{\alpha\gamma_{0}}{2}[-H_{K}sin2(\phi-\chi)-Q{\pi}H_{D}sin(\phi-\chi)+{\pi}H_{x}sin\phi-{\pi}H_{FL}cos\phi] \\ 
	& +Q\gamma_{0}[H_{pin}(q)+Q\frac{\pi}{2}H_{DL}cos\phi],
\end{split}
\end{equation}
\begin{equation} \label{1DM_3}
\begin{split}
  \alpha\frac{\pi^{2}}{12\gamma_{0}}[tan^{2}\chi+(\frac{L_{y}}{\pi\Delta{cos\chi}})^{2}]\frac{dq}{dt} & =-[\frac{2K_{eff}}{\mu_{0}M_{s}}+Q\frac{\pi}{2}H_{D}cos(\phi-\chi)+H_{K}cos^{2}(\phi-\chi) \\ 
	& -\frac{\pi}{2}H_{x}cos\phi-\frac{\pi}{2}H_{FL}sin\phi]tan\chi-Q\frac{\pi}{2}H_{DMI}sin(\phi-\chi) \\ 
	& -\frac{\pi}{2}H_{K}sin2(\phi-\chi),
\end{split}
\end{equation}
where $\Delta=\Delta_{0}/(\sqrt{1-h^{2}}\mp{hcos(\pm{h})})$, with $\Delta_{0}=\sqrt{\frac{A_{ex}}{K_{eff}}}$ ($A_{ex}=1.6\times10^{-11}$~J/m \cite{thiaville2012dynamics,hrabec2014measuring}) being the DW parameter at rest, $h=H_{x}/H_{K_{eff}}$, $K_{eff}=\frac{K_{i}}{t_{FM}}-\frac{\mu_{0}{M_{s}}^{2}}{2}$ and $H_{K_{eff}}=2K_{eff}/\mu_{0}M_{s}$. $K_{i}=2.2$~mJ/m$^{2}$ is the PMA anisotropy constant, $t_{FM}=t_{Co}$ the ferromagnetic layer thickness and $M_{s}=1.4\times10^{6}$~A/m the saturation magnetization \cite{kim2015improvement}. $H_{K}=N_{x}M_{s}$ is the DW shape anisotropy field with $N_{x}=t_{FM}Log(2)/(\pi\Delta)$ being the demagnetization field factor \cite{martinez2014current}. A large Gilbert damping parameter $\alpha=0.1$ is used as taken from literature \cite{ryu2014chiral}. $Q=+1$/$Q=-1$ for the $\uparrow\downarrow$/$\downarrow\uparrow$ configuration of DW. The DMI is modeled as an effective field along the $x$-axis with its amplitude given by $H_{D}={D}/\mu_{0}M_{s}\Delta$, where $D$ is the DMI parameter \cite{thiaville2012dynamics}. $H_{x}$ is the applied longitudinal field, $H=H_{pin}(q)$ is the pinning field. The spatially-dependent pinning field accounts for local imperfections (such as edge and surface roughness or defects), and can be derived from an effective spacial-dependent pinning potential, $V_{pin}(q)$, as $H_{pin}(q)=-1/(2\mu_{0}M_{s}L_{y}t_{FM})(\partial{V_{pin}}/\partial{q})$ ($L_{y}$ is the width of the magnetic wire). A periodic potential is employed to describe the experimental results, $V_{pin}(q)=V_{0}sin⁡(\pi{q}/p)$, where $V_{0}=7$x$10^{-19}$~J is the energy barrier of the pinning potential and $p=\Delta$ is its periodicity. Eqs. \ref{1DM_1}, \ref{1DM_2}, and \ref{1DM_3} are numerically solved by means of a 4th Runge-Kutta algorithm with a time step of 0.1~ps over a temporal window of 100~ns.
\section*{Appendix B: Role of the field-like torque in the DW motion process}
\begin{figure*}[htbp]
	\centering
	  \subfigure[]
		  {\includegraphics[width=75mm]{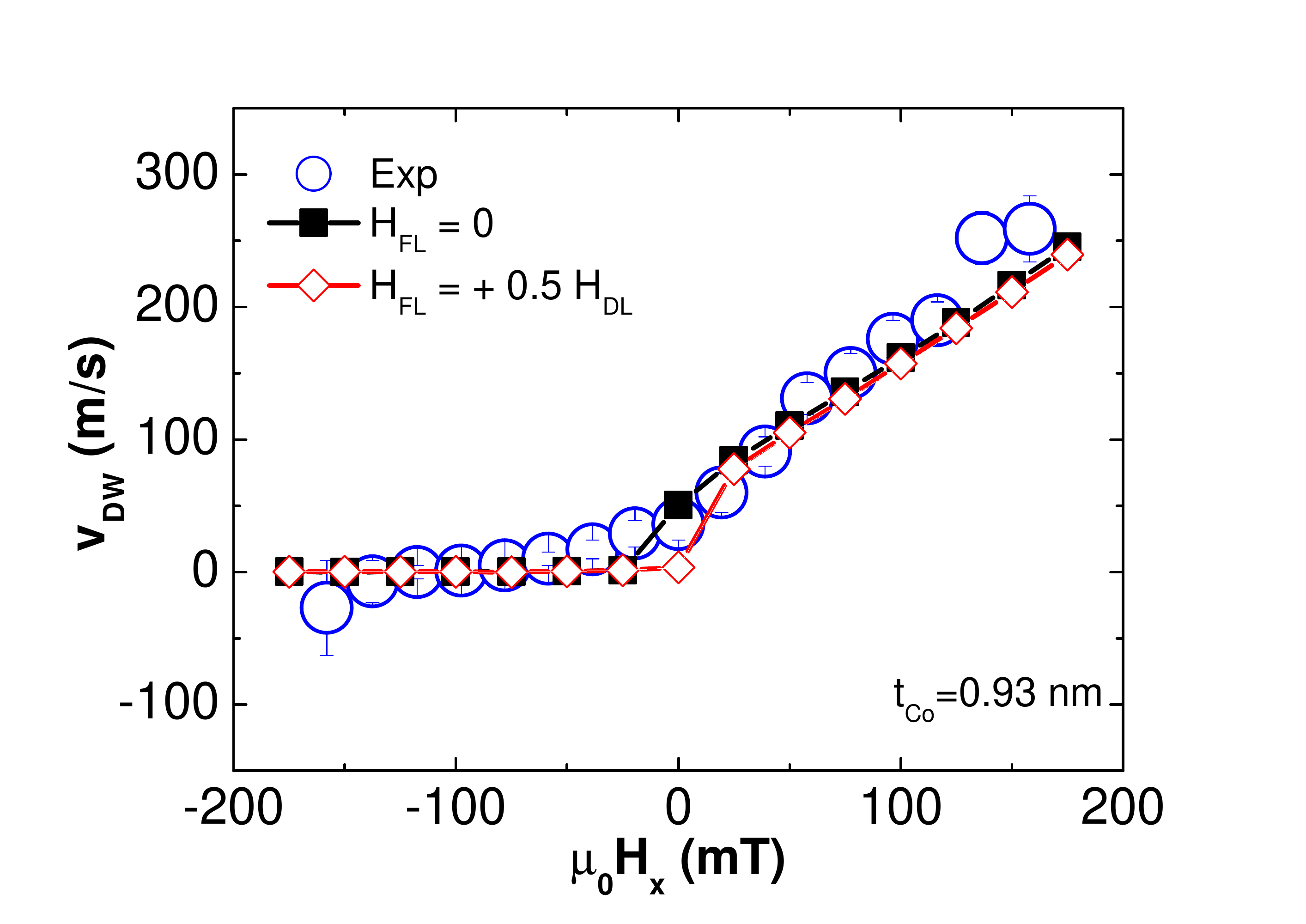}
		     \label{fig_1DM_posFL_D1}}
		\subfigure[]
		  {\includegraphics[width=75mm]{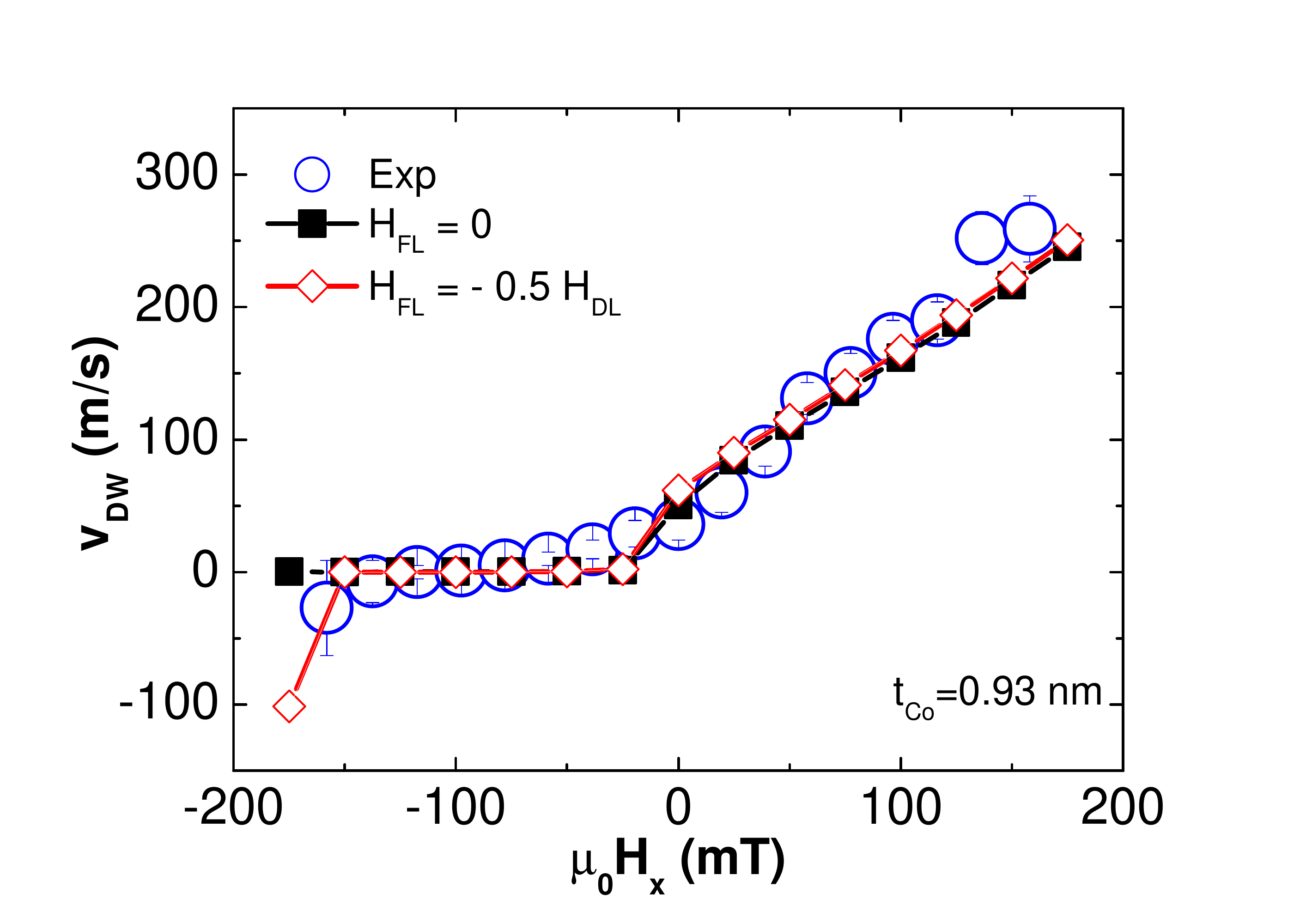}
		     \label{fig_1DM_negFL_D1}}
		\subfigure[]
		  {\includegraphics[width=75mm]{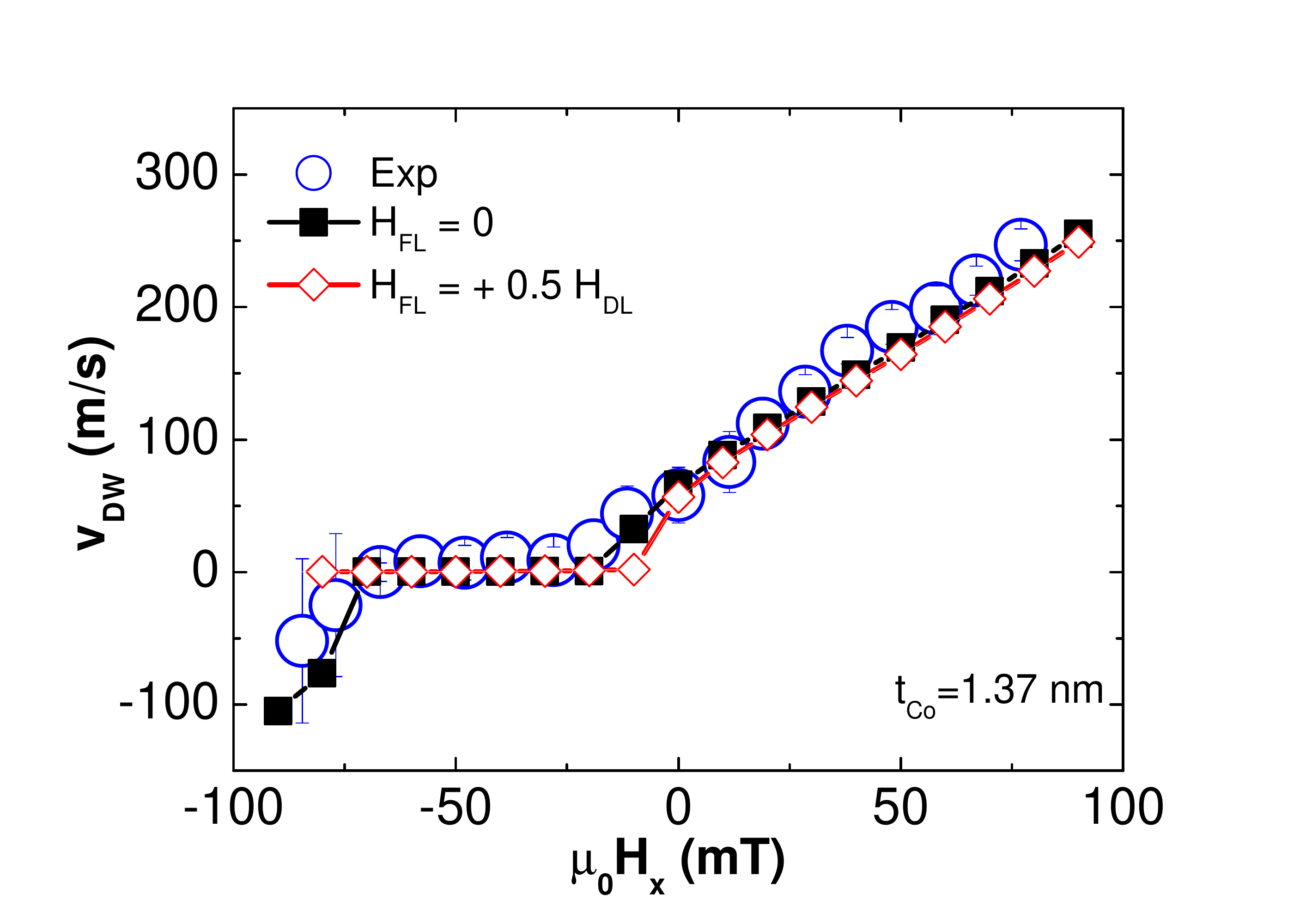}
		     \label{fig_1DM_posFL_D4}}
		\subfigure[]
		  {\includegraphics[width=75mm]{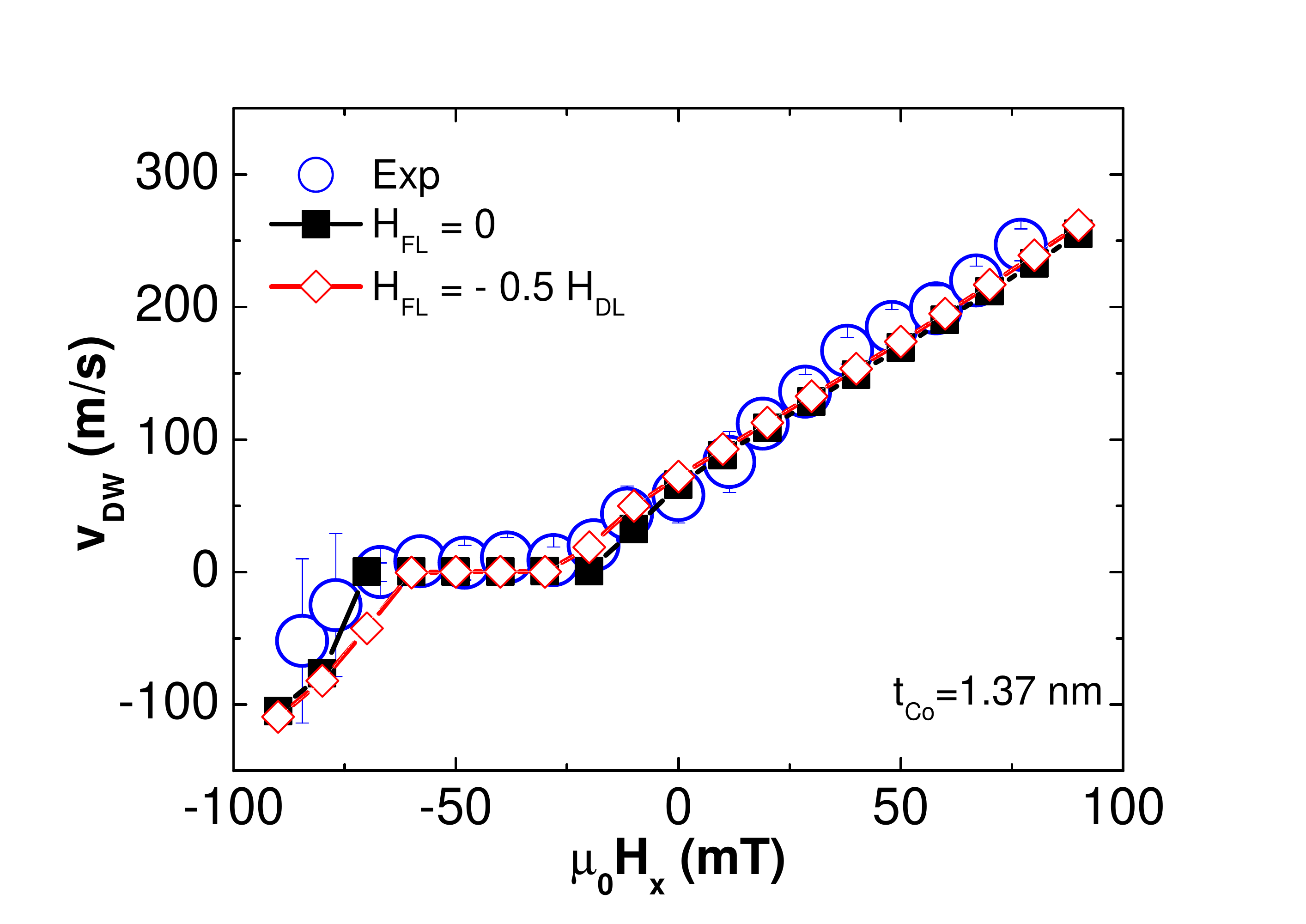}
		     \label{fig_1DM_negFL_D4}}
	\caption{Role of the FL-SOT in the definition of the final DW velocity. The blue open dots represent the experimental data for the $\downarrow\uparrow$-DW in the case of $j_{a}>0$, the solid squares are the 1DM calculation data points for $H_{FL}=0$, the open diamonds are the 1DM calculation data points for $H_{FL}=\pm{0.5H_{DL}}$. (a) and (b) refer to the device with $t_{Co}=0.93$~nm. (c) and (d) refer to the device with $t_{Co}=1.37$~nm.}
	\label{fig_1DM_FL-SOT}
\end{figure*}

The analysis based on the 1DM calculations allows us to gauge the role of the FL-torque on the DW motion. In Fig. \ref{fig_1DM_FL-SOT} the experimental results (open dots) for the device with $t_{Co}=0.93$~nm (a) and (b), and $t_{Co}=1.37$~nm (c) and (d) are reported together with several fitting curves obtained considering different $H_{FL}$ amplitudes (different $\eta$ values) but always the same $H_{DL}$ used to generate the fitting curves in Fig. \ref{fig_vDW-Hx}. In Fig. \ref{fig_1DM_posFL_D1} and \ref{fig_1DM_posFL_D4} the case $H_{FL}=0$ (solid squares) is compared with the case $H_{FL}=+0.5H_{DL}$ (open diamonds). While, in Fig. \ref{fig_1DM_negFL_D1} and \ref{fig_1DM_negFL_D4} the comparison is between $H_{FL}=0$ (solid squares) and ${FL}=-0.5H_{DL}$ (open diamonds).\\On the one hand, as visible in every graph reported in Fig. \ref{fig_1DM_FL-SOT}, the different fitting curves almost overlap each other. No net difference is visible between the curves for the high DW velocity regime, demonstrating the negligible role played by the FL-torque in the definition of the final DW velocity. Both fitting curves manage to reproduce the experimental data points with the same accuracy. This supports the initial choice of not including any FL-torque in the extraction procedure of the effective SHA. On the other hand, some differences between the fitting curves can be observed at the pinning regions. The curves obtained by including a finite $H_{FL}$ with different sign predict a different size of the pinning range (compare between Fig. \ref{fig_1DM_posFL_D1} and \ref{fig_1DM_negFL_D1}, and between Fig. \ref{fig_1DM_posFL_D4} and \ref{fig_1DM_negFL_D4}). This seems to suggest that the FL-torque can play a role in the depinning process of the DW, facilitating the DW depinning or making it harder, according to its sign with respect to the DL-torque.      
\newpage

\end{document}